\documentclass[conference,10pt]{IEEEtran}

\usepackage{epstopdf}
\usepackage{graphicx}

\usepackage{setspace}

\usepackage[utf8x]{inputenc}
\usepackage[QX,T1]{fontenc} 
\usepackage{xspace}
\usepackage{balance}
\usepackage{tikz}
\usepackage{url}
\usepackage{mdwlist}
\usepackage{subfigure}
\usepackage{mdwlist}
\usepackage{bm}
\usepackage{booktabs}
\usepackage{multirow}
\usepackage{hhline}
\usepackage[ruled,linesnumbered]{algorithm2e}
\usepackage{algorithmic}

\usepackage[font=small,labelfont=bf,skip=0pt]{caption}
\usepackage{rotating}
\usepackage{array}
\usepackage{caption}

\usepackage{hyperref}

\usepackage{color}   %for hilite command
\newcommand{\hide}[1]{}

\newcommand{\OffComm}{Offensive Community\xspace}
\newcommand{\OffCommShort}{Offensive Comm.\xspace}
\newcommand{\HTS}{hack this site\xspace}
\newcommand{\WS}{Wilders Security\xspace}

\newcommand{\Ash}{Ashiyane\xspace}

\newcommand{\WordFreq}{Words-Frequency\xspace}
\newcommand{\Mixed}{Combined\xspace}
\newcommand{\CoCluster}{Co-Clustered\xspace}

\newcommand{\reminder}[1]{{\textcolor{red}{ #1}}}
\newcommand{\mrem}[1]{{\textcolor{blue}{ #1}}}    %% Michalis comments

\newcommand{\myalg}{InferIP\xspace}

\newcommand{\Malicious}{Contributing\xspace}

\newcommand{\lfsets}{{\it latent feature sets \xspace}}
\newcommand{\blfsets}{{latent feature sets\xspace}}

\newcommand{\keyproblem}{Key Question\xspace}  %Michalis to replace Problem 1

\begin{document}

%%%%%%%%%%%%%%%%%%%% copy right
%\IEEEoverridecommandlockouts
%\IEEEpubid{
%\parbox{\columnwidth}{\vspace{-4\baselineskip}
%Permission	to	make	digital	or	hard	copies	of	all	or	part	of	
%this	work	for	personal	or	classroom	use	is	granted	without	fee	provided	%that	copies	are	not	made	or	
%distributed	for	profit	or	commercial	advantage	and	that	copies	bear	this	%notice	and	the	full	citation	on	the	
%first	page.	Copyrights	for	components	of	this	work	owned	by	others	than	%ACM	must	be	honored.	
%Abstracting	with	credit	is	permitted.	To	copy	otherwise,	or	republish,	5to	post	on	servers	or	to	redistribute	
%to	lists,	requires	prior	specific	permission	and/or	a	fee.	Request	%permissions	from	
%\href{mailto:permissions@acm.org}{permissions@acm.org}.\hfill\vspace{-0.8\baselineskip}\%\
%\begin{spacing}{1.2}
%\small\textit{ASONAM	'17},	July	31	- August	03,	2017,	Sydney,	Australia	%\\
%\copyright\space2017	Association	for	Computing	Machinery.	\\
%ACM	ISBN	978-1-4503-4993-2/17/07?/\$15.00	\\
%\url{http://dx.doi.org/10.1145/3110025.3110055}	
%\end{spacing}
%\hfill}
%\hspace{0.9\columnsep}\makebox[\columnwidth]{\hfill}}
%\IEEEpubidadjcol

% --- Author Metadata here ---
%\conferenceinfo{author metadata}{here}
%\CopyrightYear{2007} % Allows default copyright year (20XX) to be over-ridden - IF NEED BE.
%\crdata{0-12345-67-8/90/01}  % Allows default copyright data (0-89791-88-6/97/05) to be over-ridden - IF NEED BE.
% --- End of Author Metadata ---

\title{Mining actionable information from security forums: the case of malicious IP addresses}

\author{
\IEEEauthorblockN{Joobin Gharibshah\IEEEauthorrefmark{1},
Tai Ching Li\IEEEauthorrefmark{1}, Andre Castro\IEEEauthorrefmark{1},\\Konstantinos Pelechrinis\IEEEauthorrefmark{2},
Evangelos E. Papalexakis\IEEEauthorrefmark{1}  and Michalis Faloutsos\IEEEauthorrefmark{1}}
\IEEEauthorblockA{\IEEEauthorrefmark{1}
University of California - Riverside, CA\\
%900 University Ave, Riverside, California 92557\\
Email: \{jghar002,tli010,acast050,epapalex,michalis\}@cs.ucr.edu}\IEEEauthorrefmark{2}
School of Information Sciences, University of Pittsburgh, Pittsburgh, PA\\
Email: kpele@pitt.edu}

\maketitle

%\vspace{-0.1in}
\begin{abstract}

The goal of this work is to systematically extract information from  hacker forums, whose information would be in general described as unstructured: 
the text of a post is not necessarily following any writing rules.
By contrast, many security initiatives and commercial entities are harnessing the readily public information, but they seem to focus on structured sources of information.
%, which attract both security experts and hackers with debatable intentions.
%The challenge is that these forums have: (a) undefined identities and roles among its users,  unlike other social media, and (b) unstructured information, unlike say a commercial security report. 
Here, we focus on the problem of identifying malicious IP addresses, among the IP addresses which are reported in the forums. 
%While IP blacklists exist these can be slow to be updated. 
We  develop a method  to automate the identification of malicious IP addresses with the design goal of being independent of external sources.
    A key novelty is that we use a matrix decomposition method to extract latent features of  the {\bf behavioral information} of the users, which we combine with textual information from the related posts. 
    A key design feature of our technique is that it can be readily applied to different language forums, since it does not require a sophisticated Natural Language Processing approach. 
    In particular, our solution only needs a small number of keywords in the new language plus the user's behavior captured by specific features.
We also develop a tool to automate the data collection from security forums. Using our tool, we collect approximately 600K posts from 3 different forums.
 Our method exhibits high classification accuracy, while the precision of identifying malicious IP in post is  greater than 88\% in all three forums.
 We argue that our method can provide singificantly more information: 
 %Furthermore, by applying our method,
 we find up to 3 times more potentially malicious IP address  compared to the reference blacklist VirusTotal.
  %
\iffalse
As our key contribution, we develop a systematic approach for extracting useful information from hacker forums. A key novelty is that we use user profiles and behavioral patterns to help us identify and refine information that we could not get from text analysis alone.
%We develop a forum-scraping tool, and 
We collect a wealth of data from 4 different security forums. %\mrem{XXX posts from Y security forums}.
The contribution of our work is twofold; (a) we develop a method to automatically identify through the forums malicious IP addresses 
%that are being reported in the forums 
often \mrem{weeks and sometimes months} before reported on blacklists, while (b) we also propose a systematic method to classify users into behavioral groups. 
We further showcase how this information can inform knowledge extraction from the forums.
\fi
As the cyber-wars are becoming more intense, having early accesses to useful information becomes more imperative to remove the hackers first-move advantage, and our work is a solid step towards this direction.

%Online discussion groups in the form of forums have been one of the first platforms of online social networking. 
%These platforms are usually centered around a common theme of interest to its users.  
%In this work we focus on security-related forums, where users post questions and answers relevant to computing/networking systems. 
%Our objective is twofold; (a) obtain a better understanding on how users are using these platforms to meet their security needs, as well as (b) explore possibilities for identifying potential emerging and upcoming security issues.  
%In order to achieve the above goals we collect data from a set of security related forums, namely, Wilders Security and Offensive Community.
\end{abstract}

{\bf Keywords: Security, Online communities mining, Forums}

\section{Introduction}
\label{sec:intro}

How can we take the first-mover advantage away from hackers?
We argue that hacker forums provide information earlier
than other sources, and we should leverage these forums
in our security intelligence.
Here, we focus on a specific question. 
In particular, we want to extract as much useful information from hacker/security forums as possible in order to perform (possibly early) detection of malicious IP addresses, e.g., prior to their appearance on blacklists. 
The latter can exhibit large delays in their update and hence, new ways for labeling malicious IP addresses are needed \cite{Hang2016}.
In this study we will use the term ``hacker forums'' to describe online forums with a focus on security and system administration. 
%\mrem{Can we quote how many such forums exist, rough number of users per forum etc? A simple google search, not hours of research?}
Interestingly, we can classify these forums into categories: (a) main stream forums, like  \WS,
and (b) ``fringe'' forums, like \OffComm, where we find users with names like {\it satan911}. 
Some of these forums have been known to have hackers boast of attacks they have mounted, or sell tools for malicious purposes (think rent-a-botnet). For example, in our dataset there is a post that mentions ``I give you a second server to have your fun with. Multiple websites on this server. So let's see if anyone can actually bring down the server''. Right after that the hacker posted the IP, username and password for anyone to access the server. 
In fact, there is a \textit{show-off} section in these forums for people to broadcast their hacking ``skills''.
%These fringe forums are the entry way to the dark web, which is reportedly
%Second, we can distinguish forums based on the language they use.

%Problem: 
% To summarize, the central theme in our work is to develop efficient techniques for extracting information from a security forum with the goal of informing a security analyst. The particular problem of our study is to identify malicious entities, and more specifically malicious IP addresses. 

The overarching goal of this work is to mine the 
unstructured, user-generated content in security forums. Specifically, we focus here on collecting malicious IP addresses, which are often reported at such forums. 
We use the term security forum to 
refer to discussion forum  with a focus on 
security, system administration, and or more generally, systems-related discussions.
The users in these forums include: security professionals, hobbyists, and hackers, who go on these forums
to identify issues, discuss solutions, 
and in general exchange information.

Let us provide a few examples of how users report IP addresses, which may or may not be malicious.  Posts could talk about a benign IP address, say in configuration files, 
%a server that may not be responding or exhibiting unusual delays. 
 as in the post: 
 
 \textit{ "[T]his thing in my hosts file:   64.91.255.87 ... [is] it  correct?"}.
 
At the same time, posts could also  report compromised or malicious IP addresses,
as in the post:

\textit{ "My browser homepage has been hijacked to  http://69.50.191.51/2484/"}.

The challenge is to automatically distinguish between
the two. By doing so, we can  provide a new source of information
of malicious IP addresses directly from the affected
individuals.
Formally, we can state the  problem  as follows: 

% \vspace{0.2cm}
% {\bf \keyproblem:
%  Malicious IP Detection.} %\label{problem:sumproblem} REPLACE THIS with \keyproblem
% Given a set of posts $\mathcal{P}_F$ that may contain IP addresses and users $\mathcal{U}_F$ of a security forum $F$, as well as, the features $\Phi_p,~\forall p\in\mathcal{P}_F$ and $\Phi_u,~\forall u\in\mathcal{U}_F$ for the posts and the users respectively, what is the probability $\Pr[   M_i = 1 | B_i = 0 ]$ that a given IP $i$ is  malicious ($M_i =1$) given that it is not reported in a reference blacklist ($B_i = 0$). 
% %\reminder{How about this, "How probabilistic model can Identify if an IP $i$ is malicious given that it does not appear in a pre-defined set of IP blacklists" I think we cannot report the probabilities here}.
% \vspace{0.2cm}

\vspace{0.2cm}
{\bf \keyproblem: Malicious IP Detection.} %\label{problem:sumproblem} REPLACE THIS with \keyproblem
Given a set of posts $\mathcal{P}_F$ that may contain IP addresses and users $\mathcal{U}_F$ of a security forum $F$, as well as, the features $\Phi_p,~\forall p\in\mathcal{P}_F$ and $\Phi_u,~\forall u\in\mathcal{U}_F$ for the posts and the users respectively, 
can we determine if a given IP address $i$ is malicious or not?
% what is the probability $\Pr[   M_i = 1 | B_i = 0 ]$ that a given IP $i$ is  malicious ($M_i =1$) given that it is not reported in a reference blacklist ($B_i = 0$). 
%\reminder{How about this, "How probabilistic model can Identify if an IP $i$ is malicious given that it does not appear in a pre-defined set of IP blacklists" I think we cannot report the probabilities here}.
\vspace{0.2cm}

\iffalse
\mrem{@Kostas: I commented out your version. It sounds cooler your way, but if we want to keep the "probability" of being malicious, we should find a way to actually answer it. Our approach gives a label. If it gives a number, we should then discuss about a threshold that we use to transform the number into a label in our methodology.}
\mrem{@Michalis: Most of the classifications the output is a probability -- in fact the logistic regression is a genuine probability. Then the label is a simple quantization of the probability at 0.5 (i.e., >0.5 label 1 otherwise 0.  So we have this number.}
\fi

The set of features $\mathcal{P}_F$ includes attributes such as the text of the post, the posting user, the time of post, etc., while $\mathcal{U}_F$ includes information such as the date of a user joining the forum, the number of posts the user has made etc.
The above problem has two associated questions:
%\newline

{\bf a. Exclusivity:} How many IP Addresses can we find that are never reported by other reference sources?

%\newline

{\bf b. Early warning:} How much earlier are malicious IP Addresses reported in a forum compared to reference sources, for the IP Addresses reported by both?

    \begin{table*}[ht]
   	\normalsize
    \renewcommand{\arraystretch}{1.3}
    \centering
    \caption{Extracting useful information; Number of malicious IP Addresses found by \myalg  and not by VirusTotal. }
    \label{tab:test_result}      
	\begin{tabular}{c|c|c|c}
	\hline
    \multicolumn{2}{c}{} & \multicolumn{2}{|c}{ IP found by} \\
    \hline
    Dataset & Total IP &  Virus Total & \myalg only \\
    \hline
    \WS & 4338   & 216 & \textbf{670} \\
    \OffComm & 7850  & 339 & \textbf{617}\\
    Ashiyane & 8121  & 133 & \textbf{806} \\
    \hline
	\end{tabular}
	\end{table*}

%The goal is to identify useful security information, which we make more concrete by focusing it on identifying malicious entities, such as malicious IP addresses. \reminder{more description}
%A stepping stone is to develop techniques to: 
 
 %OLD
 %(a) look at the right places with the right filters,
 %and (b) have context to interpret the information properly. A simplistic example is that the expertise of %a user can provide different weight on the post that identifies a problem.
 
 %Previous work:
 Most previous studies in this area have focused on  structured information sources, such as security reports,
 or malware databases. In fact, many efforts  focus on addressing security problems using knowledge obtained from the web, as well as social and information networks. These efforts are mainly focused on analyzing structured sources (e.g., \cite{Iannacone2015}).  
 However, studies assessing the usefulness of (unstructured) information in online forums have only recently emerged (e.g., \cite{Az1}).  
 These studies are rather exploratory and provide evidence of the usefulness of the data in the forums, but do not provide a systematic 
 methodology or ready-to-use tools, which is the goal of our work.
 We discuss existing literature in more detail later in  section~\ref{sec:related}.
 
 The motivation of our work is to provide more information to security analysts and systems. We want  to enhance and complement, but not replace, existing efforts for detecting malicious IP Addresses. 
    For instance, many IP blacklists enlist an IP as malicious after a number of reports above a pre-defined threshold have been made for the specific address. 
  Depending on the threshold and the reactivity of the affected users/systems, this might take several days, weeks or months. 
  Therefore, a system, like the one proposed here, can identify and point to malicious IP address to blacklist services  and firewalls.
  
  %if $\Pr[M_i = 1 |  B_i = 0] > \theta$, where $\theta$ is a pre-defined threshold. 
  %\mrem{Let's remove the next phrase:  Furthermore, while in this work we do not focus necessarily on the timely detection of malicious IP addresses, a system based on the analysis of security forums can certainly drive this as well. 
  %}
  
 %There has been some work in the area of analysing security domain on web and social networks.
 %We can identify the following categories of related work:
 %(a) efforts that analyze structured sources of security information,\cite{Jones2015,Iannacone2015} \\
 %and (b) exploratory efforts assessing the usefulness of information in online forums
 %\cite{Az1} \reminder{more citation}. The efforts in the latter group do not seem to have lead to automated and massively scalable techniques, which is our goal here. We discuss previous work in section \ref{sec:related}.

 %Contributions

  We propose \myalg,  a systematic approach for  identifying malicious IP Addresses among the IP addresses, which are mentioned in  security forums.
  %A  goal is to extract malicious IP Addresses accurately and scalably from a large numbers of forums.
    A key novelty is that we use  the  behavioral information of the users, in addition to the textual information from the related posts. Specifically, we customize and use a Sparse Matrix Regression method on this expanded set of features.
    
    This paper presents an extension of our previous work \cite{Joobin2017}. Here, we add some spatiotempral and behavioral analysis to extract the characteristics of the identified IP addresses and the users who used these IP address in their posts. Moreover, we investigate the ability of the proposed method to provide early warning regarding malicious IP addresses. 
    
    By design, our framework is applicable to forums in different languages as it relies only on the behavioral patterns of users and simple word counts,  and not a complex language-specific  Natural Language Processing technique. 
    %We design a method to identify malicious  IP  addresses among those reported in  the forums.
    From a technical point of view the challenge in designing a solution to our \keyproblem is most IP Addresses mentioned in these forums are not malicious. We show that our system can add a significant number of previously unreported IP address to existing blacklist services. 
    Finally, as an engineering contribution, we develop a customizable tool to facilitate the crawling of forums, which we discuss in the next section.
  
  Our results can be summarized into the following points:

  {\bf a. Our method exhibits precision and recall greater than 88\% and 85\% respectively, and an accuracy over \textit{malicious} class above 86\%} in the 10-fold cross validation tests we conducted for the three different forums.
  In partially answering our \keyproblem, if our method labels a currently non-blacklisted IP as malicious, there is a high chance that it is malicious, given our high precision.
  
%   We use VirusTotal \cite{virustotal} as our reference blacklist IP addresses,
  
  %{\bf b. Using the behavioral properties increases the accuracy} by 7\% compared to using only textual information. 
  
  {\bf b. Our method  identifies  three times more malicious IP Addresses}  compared to VirusTotal \cite{virustotal} a widely used aggregator of 60 blacklists of IP addresses. Across our three forums, we find more than 2000 potential malicious IP Addresses that were never reported by VirusTotal.
  
  %{ \color{red}

 {\bf c. Our method  identifies  more than half of the IP addresses at least 3 month earlier than VirusTotal}. We study the malicious IP addresses that are identified by both VirusTotal and \myalg. We find 53\%, 71\% and 62\% of these IP addreses in \WS, \OffComm and \Ash  respectively  at least 3 months earlier than they were reported in VirusTotal.

 {\bf d. The number of reported malicious IP addresses has increased by a factor 8 in 4 years.} We find that the number of malicious IP addresses has increased from roughly 100 in 2011 and 2012 to more than 800 in 2016. This could be attributed to
 either an increase in the user base, an increase in
 the number of attacks, or a combination of the two.
 
%  {\bf d. Our method identifies a diverse range of IP addresses located in different continents.} In spite of dynamic distribution of the malicious IP addresses' geo-location over years and differences in distribution in \myalg and VirusTotal methods' results, on average over years, both methods show same order in geo-location distribution (i.e. Table \ref{tab:AvgDist}).
% %}

  \iffalse
  {\bf c. Leveraging user profile and behaviors.} We develop an unsupervised technique to find groups of users based on their behavioral profiles. We start with a set of features that describe the user behavior, and by doing \mrem{co-clustering}, we identify a single digit number of profiles.
  Using these profiles, we show that we can know: (a) where to look, and (b) how to interpret useful information.
  \mrem{provide concrete examples - at the very least we can describe the groups and say that ~90\% of users seem to be "consumers" posting very little and mostly on a thread that they start}
  \fi

  %The rest of the paper is organized as follows: 
  %Section \ref{sec:data} briefly describes our datasets  and some basic statistics. 
  %Section \ref{sec:ipmodel} presents the design and evaluation of our model for identifying malicious IP addresses.  
  %Finally, Section \ref{sec:conclusions} concludes our work and outlines our future directions. 

\section{Data Collection and Basic Properties}
\label{sec:data}

We have collected data from three different forums relevant to our study; (i) \WS \cite{wilders}, (ii) \OffComm \cite{offcomm}, (iii) \Ash \cite{ashiyane}.  
The first two forums are mainly written in English, while the last forum is an Iranian forum, in Farsi\footnote{Our software and datasets will be made available at: \url{https://github.com/hackerchater/}
}.

{\bf Our data collection tool.} We develop a customizable universal tool to make the  crawling forums easier. The challenge here is that each forum has its own format and layout. Our tool requires only a custom configuration file, before crawling a new forum. In configuration file, we specify entities in the forum which are needed such as user ID, post's date, post's content and etc by XML Path Language known as \textit{Xpath}. Leveraging our current configuration files, the task of crawling a new forum is simplified significantly. 
  Using our crawler, we collect data from three forums, two English and one in Farsi for a total number of more than 30K users and 600K posts.

  We use VirusTotal \cite{virustotal} as our reference blacklist IP addresses, since it is an aggregator, and combines the information from over 70 other blacklists and resources. VirusTotal is free to end users for non-commercial use and is a private API to query the services in the rate of more than 4000 IP addresses per minutes. It is provided upon requests for academic purposes.

We provide some basic statistics for our three forums  in Table \ref{tab:forums}. 
\OffComm and \Ash are two fringe forums in different languages. In these forums there is a section where people openly boast about their achievement in hacking. They share their ideas and {\em tutorials} on how to break into vulnerable networks. On the other hand, \WS as a mainstream forum is mostly used to protect non-experts against attacks such as browser hijacking, and provide solutions for their security problems. %\reminder{breifly I have discuss about these things at beginnig of the introduction. Don't you think it is redundant }

For completeness, we present some of the terms we use here. A user is defined by a login name registered with the site.
The term post refers to a single unit of content generated by a user. A thread refers to a collection of posts that are replies to a given initiating post.

\begin{figure*}[!htb]
\begin{minipage}{0.32\linewidth}
  \includegraphics[width=\linewidth]{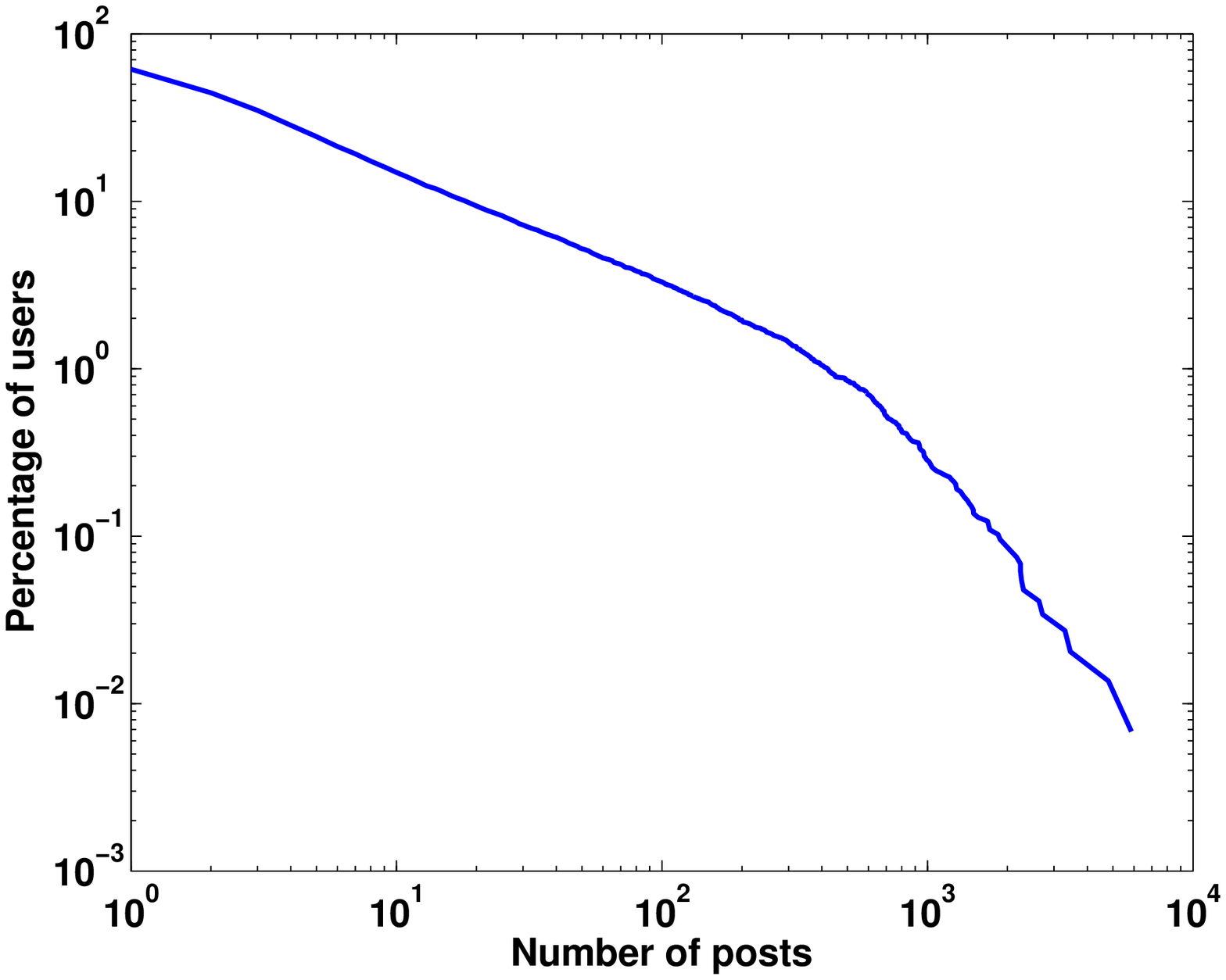}
  \centering
  { (a) \WS.}
\end{minipage}
\begin{minipage}{0.32\linewidth}
  \includegraphics[width=\linewidth]{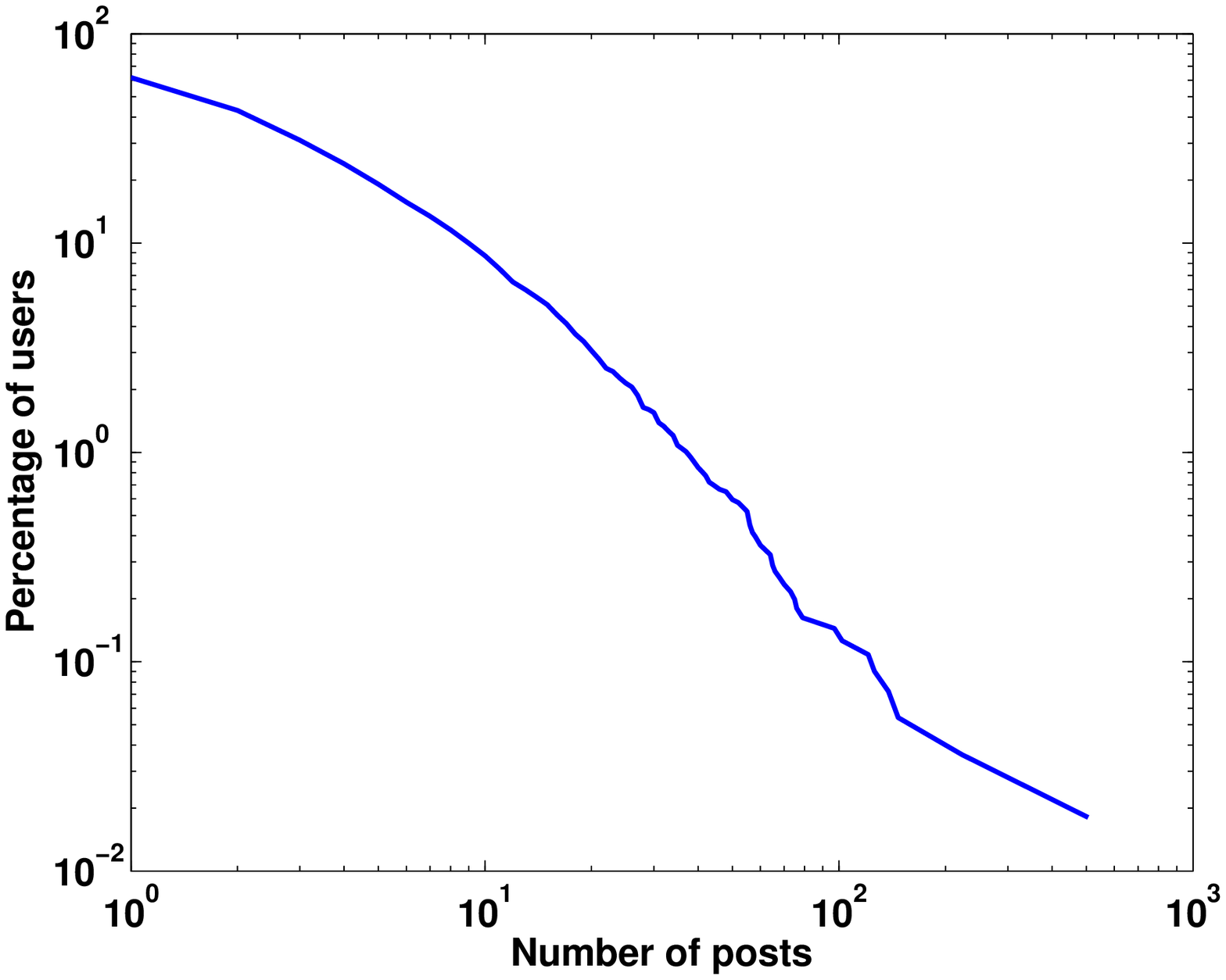}
  \centering
  {(b) \OffComm.}
\end{minipage}
\begin{minipage}{0.32\linewidth}%
  \includegraphics[width=\linewidth]{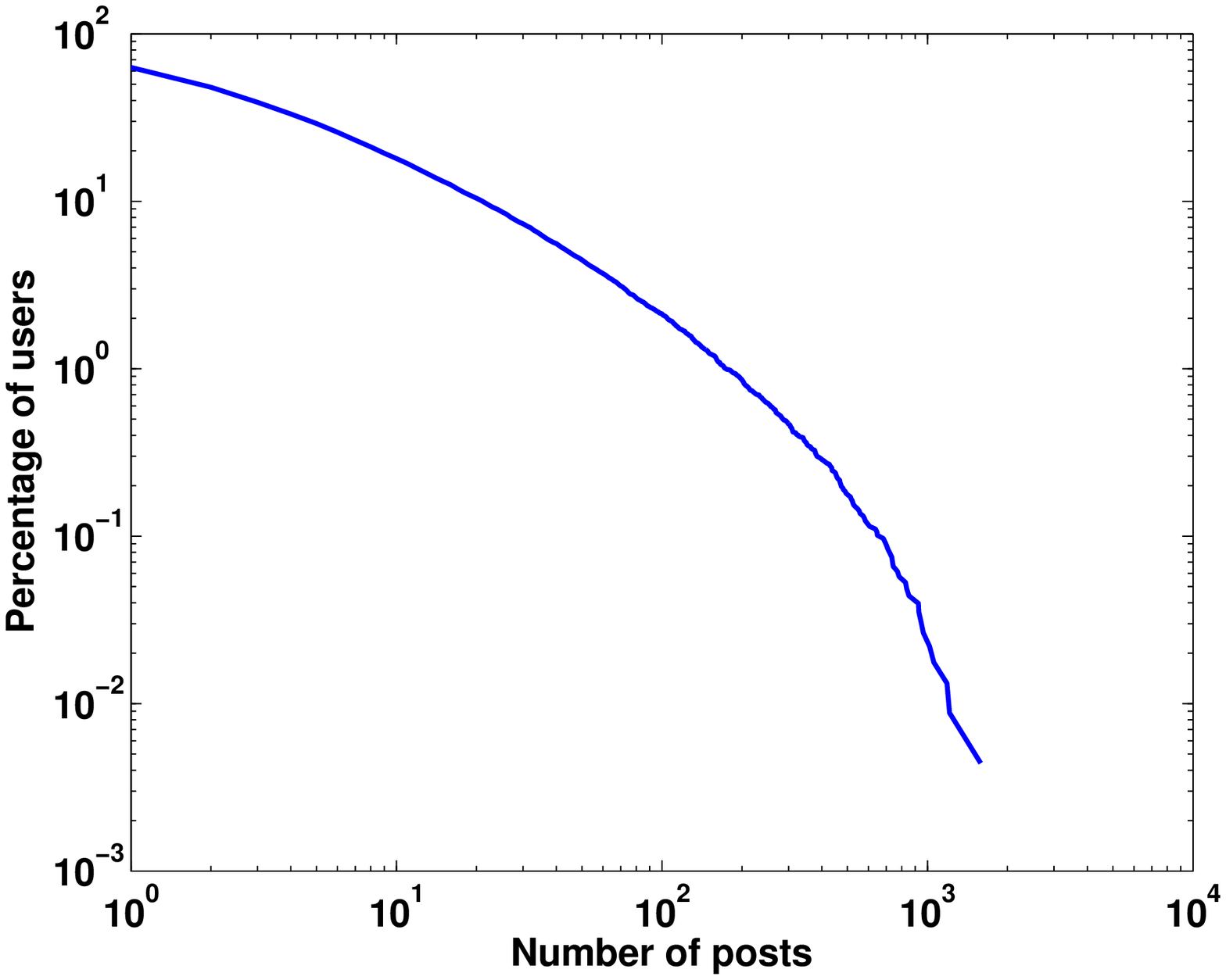}
  \centering
  {(c) \Ash}
\end{minipage}
\caption{CCDF of the number of posts per user ($\log$-$\log$ scale).}\label{fig:postperuser}

\end{figure*}

\begin{figure*}[!htb]
\begin{minipage}{0.32\linewidth}
  \includegraphics[width=\linewidth]{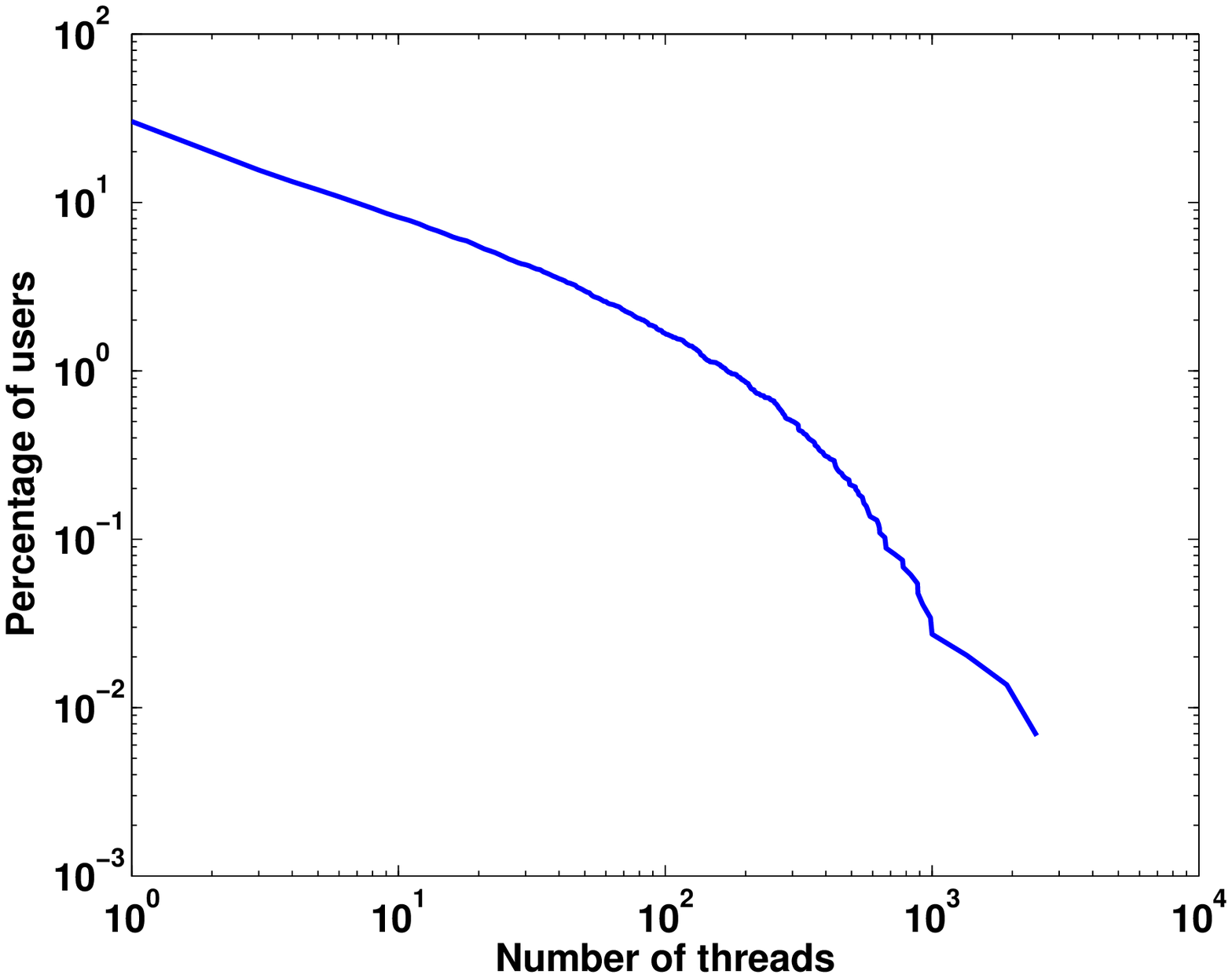}
  \centering
  {(a) \WS}
\end{minipage}
\begin{minipage}{0.32\linewidth}
  \includegraphics[width=\linewidth]{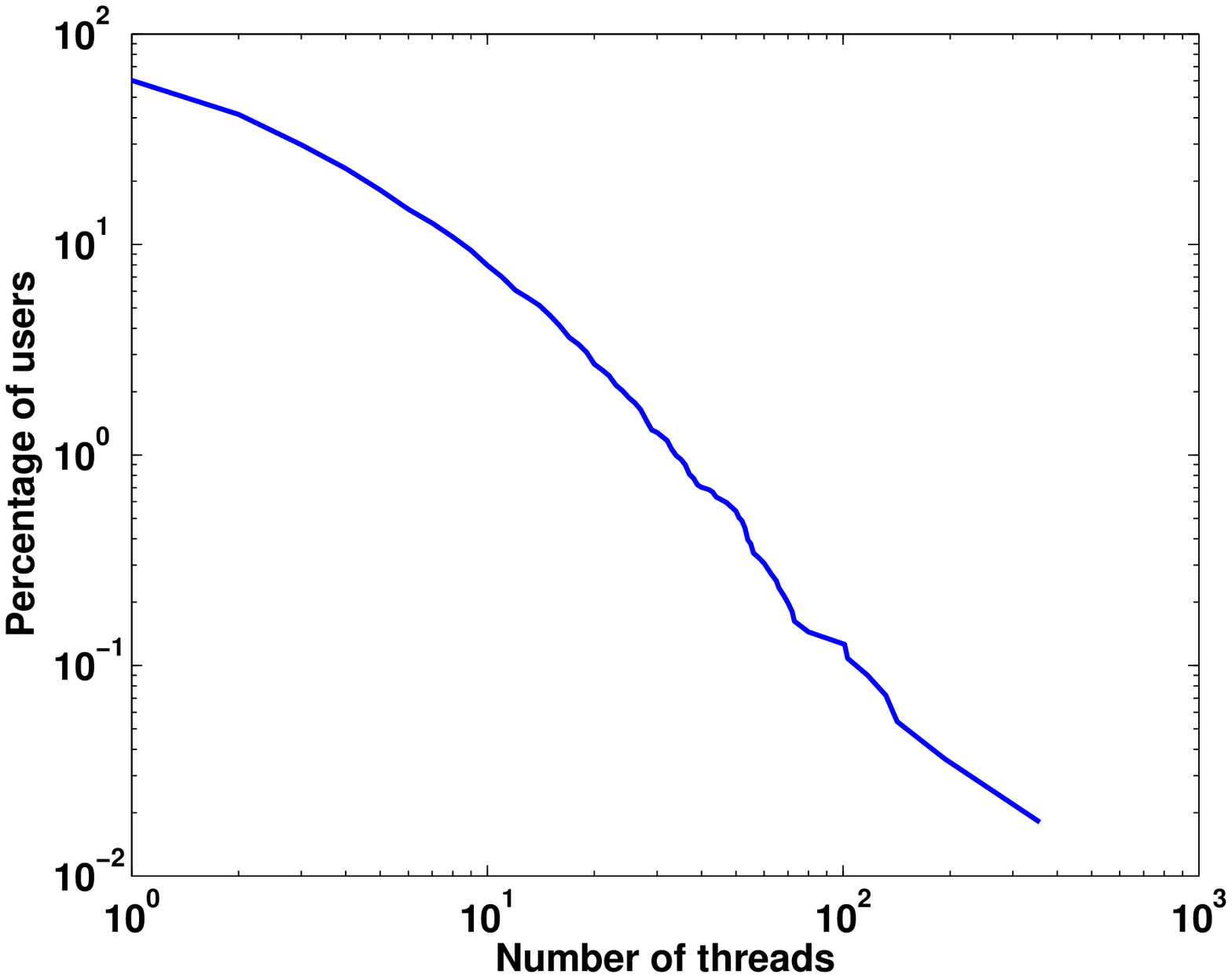}
  \centering
 {(b) \OffComm}
\end{minipage}
\begin{minipage}{0.32\linewidth}%
  \includegraphics[width=\linewidth]{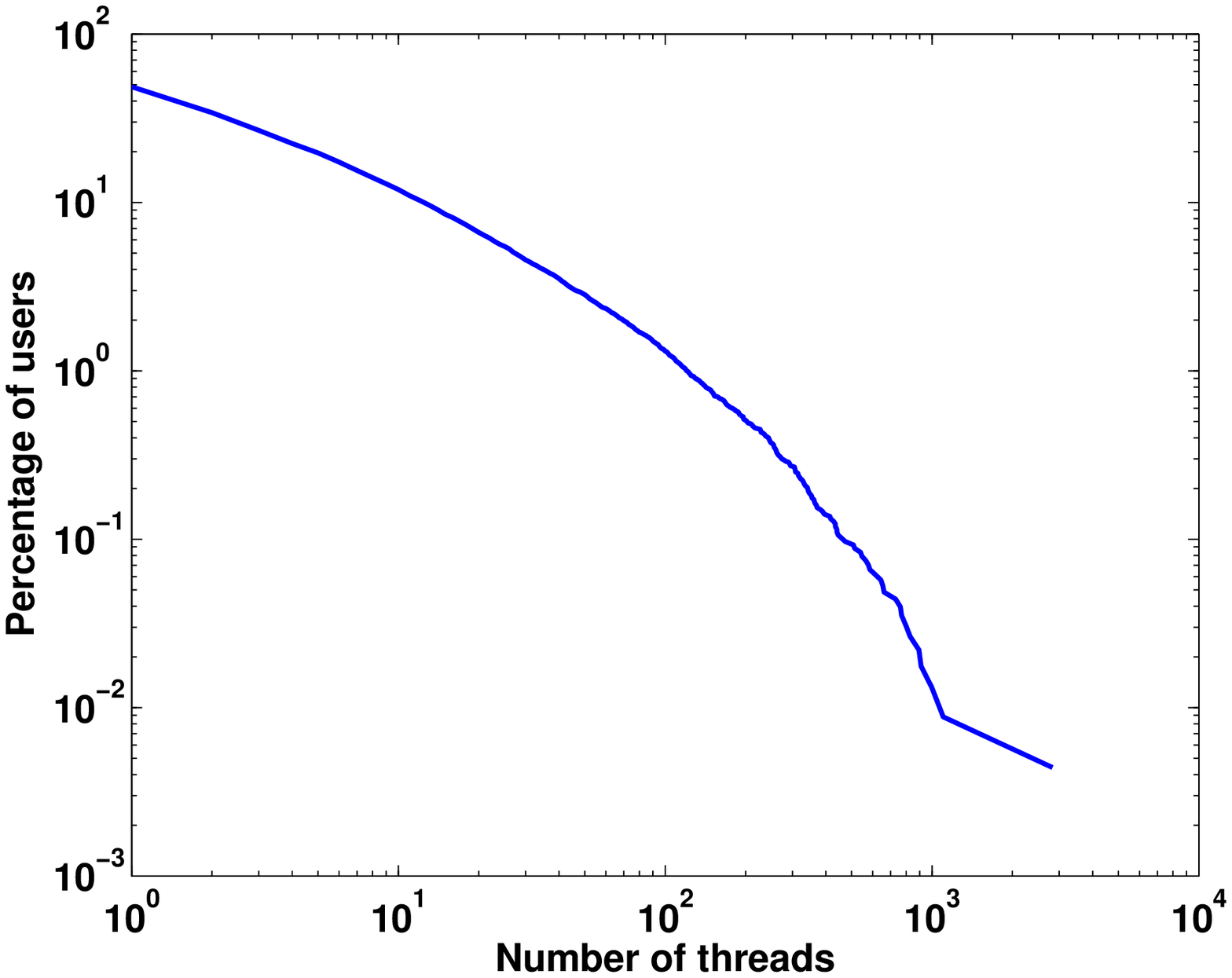}
  \centering
  {(c) \Ash}
\end{minipage}
\caption{CCDF of the number of thread per user ($\log$-$\log$ scale).}\label{fig:postperthread}
\end{figure*}

In Figures \ref{fig:postperuser} and \ref{fig:postperthread}, we present the cumulative complementary distribution function of the number of posts per user and the number of threads per users respectively.  
As we can see in all the cases the distributions are skewed, that is, most of the users contribute few posts in the forums and 
%most threads have a small number of posts. 
engage with few threads. In \WS, 
85\% of users post less than 10 posts each, while 5.2\% of the users post more than 50 posts. 
We find that 70\% of the users post in only one thread and only 8\% of the users are active in more than 10 threads.
This skewed behavior is typical for online users and communities  \cite{Devineni:2015:WTP:2808797.2808880}.
We will use features to capture aspects of both these user properties, as we will see in the next section. 

% Due to space limitations, we cannot present plots for more features that we use in our classification.

\begin{figure*}[!htb]
\begin{minipage}{0.32\linewidth}
  \includegraphics[width=\linewidth]{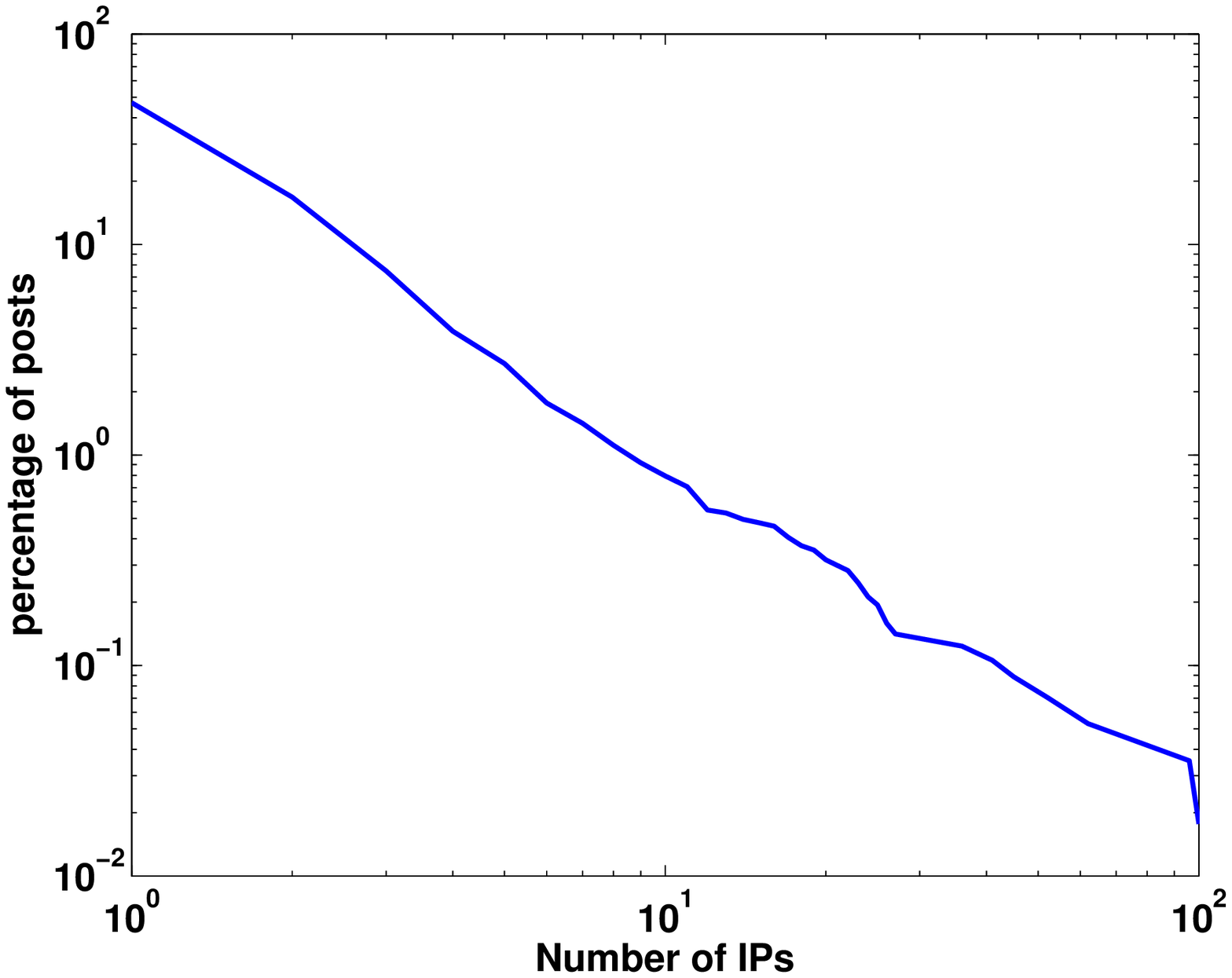}
  \centering
  {(a) \WS}\label{fig:IPPerPostWS}
\end{minipage}
\begin{minipage}{0.32\linewidth}
  \includegraphics[width=\linewidth]{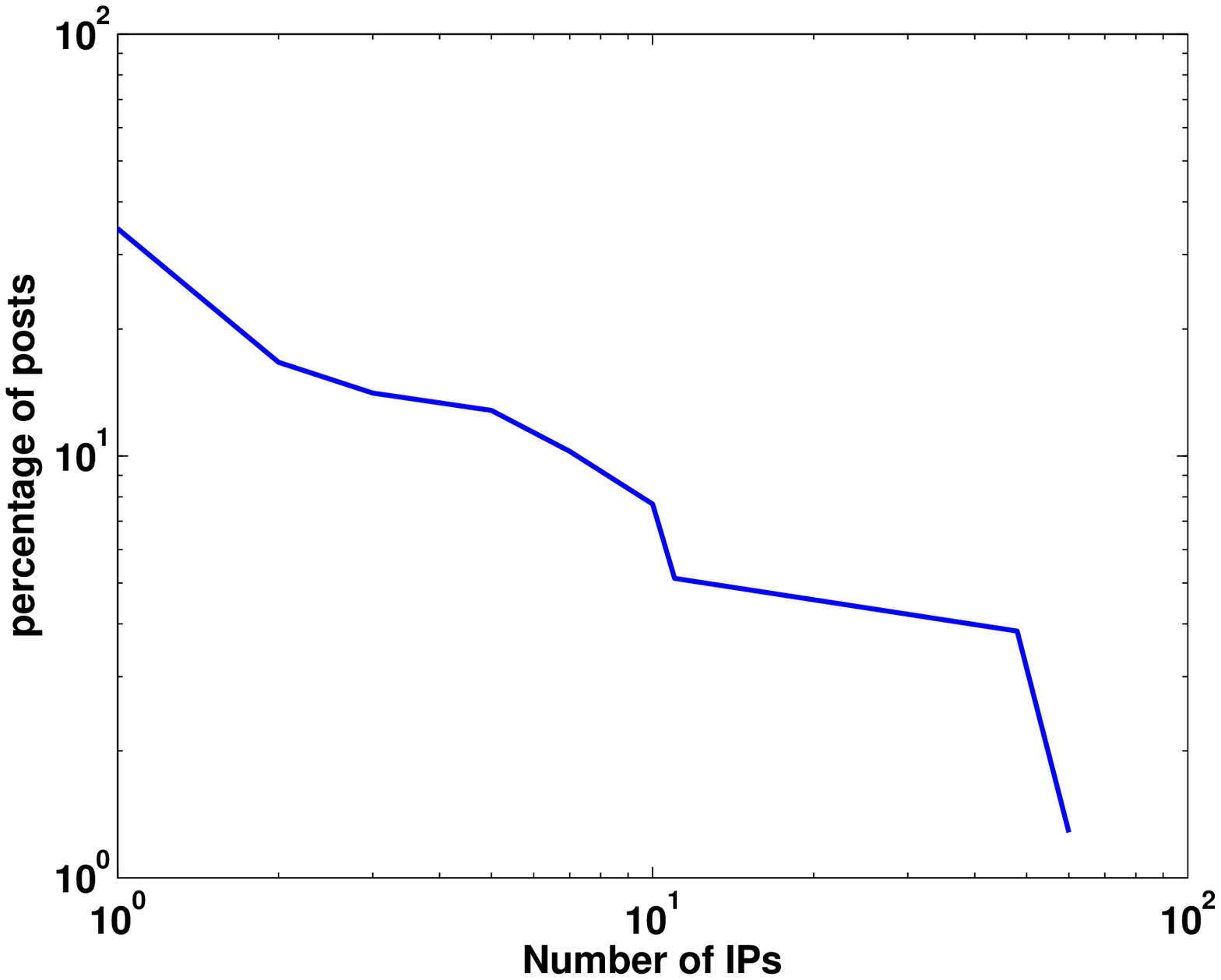}
  \centering
  {(b) \OffComm}\label{fig:IPPerPostOffComm}
\end{minipage}
\begin{minipage}{0.32\linewidth}%
  \includegraphics[width=\linewidth]{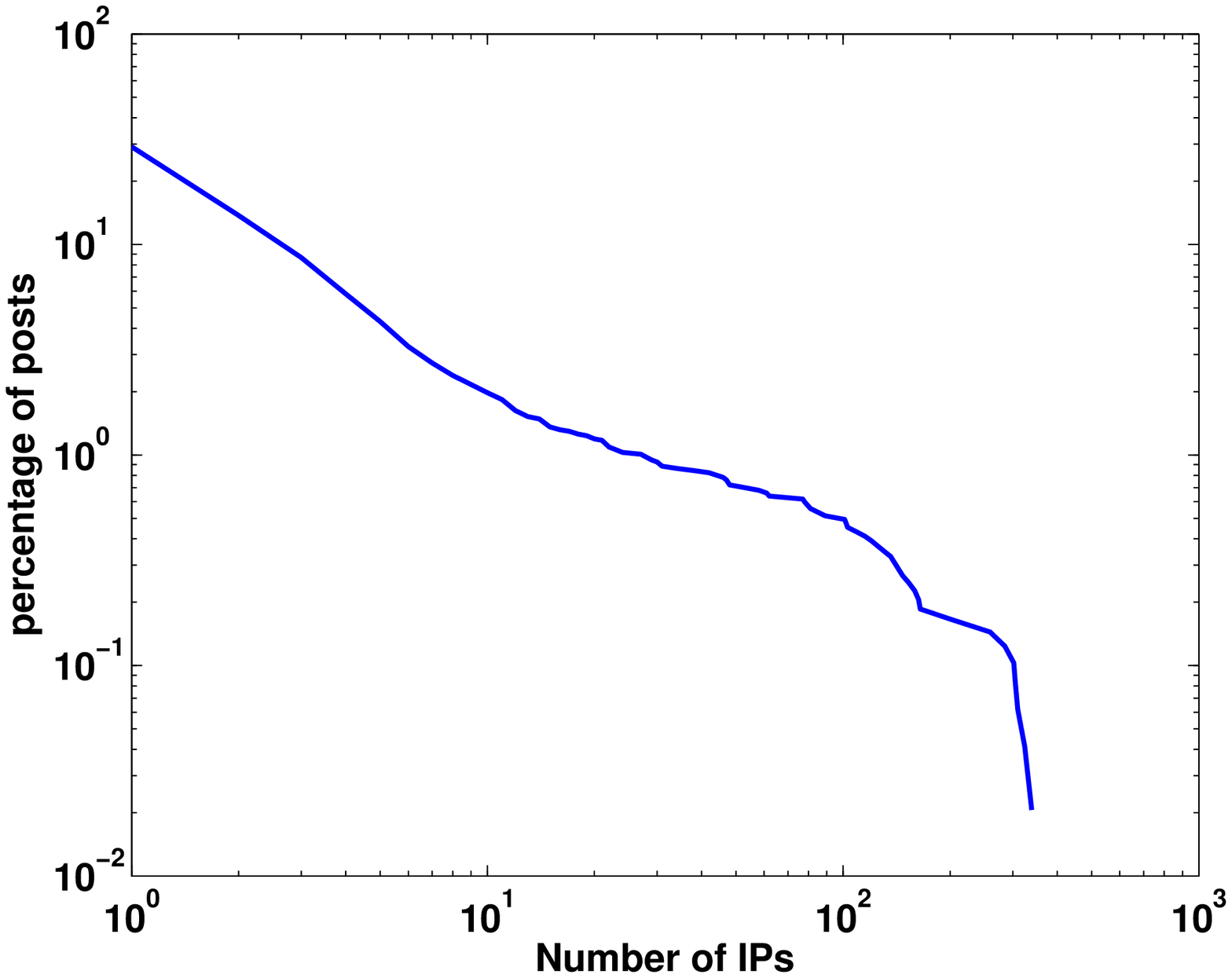}
  \centering
  {(c) \Ash}\label{fig:IPPerPostAsh}
\end{minipage}
\caption{CCDF of the number of IP addresses per post ($\log$-$\log$ scale).}\label{fig:IPPerPost}
\end{figure*}

%{ \color{red}
In Figure \ref{fig:IPPerPost}, we present the cumulative complementary distribution function of the number of IP addresses that appear in each post. The skewed distribution shows that most of the posts contain a few number of IP address. We find that  84.2\% of the posts with IP addresses in \WS and 84.1\% in \OffComm  have two or less IP addresses. In \Ash, 87.2\% of these posts contain less than two IP addresses. Interestingly, in \Ash, we find 1\% of the IP containing posts with more than 100 IP addresses. 
We investigated and we found that 
typically, these posts provide benign IP addresses of proxies servers to fellow administrators.
%}

%\vspace*{-8 mm}
\begin{table*}
\normalsize
 \centering 
 \caption {The collected forums.}
\label{tab:forums}
 \begin{tabular}{| c | c | c | c | c|} 
 \hline
 Forum & Threads & Posts & Users & Active days \\ 
 \hline
 \WS & 28661 & 302710 & 14836 & 5227 \\ 
 \hline
 \OffCommShort & 3542 & 25538 & 5549 & 1508 \\ 
 \hline
 %HackThisSite & 10135 & 68465 & 5904 & 3206 \\
 %\hline
 \Ash & 67004 & 279309 & 22698 & 4978 \\
 \hline
\end{tabular}
\end{table*}

%To solve Problem 1 %\reminder{we only have one problem isn't it better to say our problem instead?} 
{\bf Groundtruth for training and testing.}
    In order to build and evaluate, our model we need to obtain a reasonably  labeled dataset from IP addresses that appear in the posts of the security forums.  
    For that, we use the VirusTotal  service and assign malicious labels to an IP that has been reported by this service. The number of malicious IP Addresses that we have used with the corresponding posts are shown in table \ref{tab:test_result} as the IP found by VirusTotal.
    Note that the  absence of a report on VirusTotal does not necessarily mean that the IP is  benign. However, a listed IP address is most likely malicious, since VirusTotal as most blacklist sites require a high threshold of confidence for blacklisting an address.
        This way, we find in total 688 malicious IP addresses for our forums 
    as shown in Table \ref{tab:test_result}.
    
    %We also assign a \reminder{time} time stamp $t_i$ for the time when IP $i$ was reported as malicious from {\tt VirusTotal}. 
    %This will provide us with the required data for evaluating the timeliness of our detection. 
    
    Using this labeling process we have collected all the IP addresses that have appeared on our forums prior to their report on VirusTotal.  
    For building our model, we also randomly select an equal number of IP addresses that have not been reported as malicious and via manual inspection further assess their status. 
    Finally, for every security forum we have a different dataset and hence, we build a different model.

%\vspace*{-2 mm}
%\setlength{\belowcaptionskip}{-20pt}

\iffalse
\begin{figure}[!htb]
\begin{minipage}{0.32\linewidth}
 \includegraphics[width=\linewidth]{postDist_WS}
 {WildersSec.}
\end{minipage}\hfill
\begin{minipage}{0.32\linewidth}
 \includegraphics[width=\linewidth]{postDist_OffComm}
 {OffensiveCo.}
\end{minipage}\hfill
\begin{minipage}{0.32\linewidth}%
 \includegraphics[width=\linewidth]{postDist_Ash}
 {Ashiyane}
\end{minipage}
\end{figure}
\fi

%\vspace*{0 mm}

%\begin{figure}
%\subfloat[ {\tt WildersSecurity}] {\label{fig:test1}
%  \includegraphics[width=.33\linewidth]{threadDist_WS}}
%\subfloat[{\tt OffensiveCommunity}] {\label{fig:test2}
%  \includegraphics[width=.33\linewidth]{threadDist_OffComm}}
%\subfloat[{\tt Ashiyane}] {\label{fig:test3}
%  \includegraphics[width=.33\linewidth]{threadDist_Ash}}
%\caption{CCDF of the number of thread per user ($\log$-$\log$ %scale).}
%\label{fig:posts-threads}
%\end{figure}
\section{\myalg: Malicious IP Detection}
\label{sec:ipmodel}
%\textbf{Identifying upcoming security issue over IP analysis} \\*

We propose a method to identify whether an IP address within a post is malicious.  
For example, although many users report a malicious IP address, such as one that is attacking the user's network,
there are also users that will mention a benign IP address when people discuss about network tutorials like setting up \textit{Putty} or initiating a \textit{SSH} connection.

While this task is simple for a human, 
it is non-trivial to automate.
Adding to the challenge, different communities use different terminology and even different languages altogether (english and farsi in our case). 
In order to overcome these challenges, we use a diverse set of features and build a model to identify IP addresses that are potentially malicious. 

Our approach consists of four steps that each hide non-trivial novelties:

{\bf Step 1:} We consider the user behavior and extract features that profile users that post IP-reporting posts. %are more likely to report a malicious IP.

{\bf Step 2:} We extract keywords from the  posts and use information gain to identify the 100 most informative features.

{\bf Step 3:} We  identify meaningful \lfsets using an unsupervised co-clustering approach \cite{papalexakis2013k}.

{\bf Step 4:} We train a classifier using these \lfsets using 10-fold cross validation.

We describe each step in more detail.
%The key things that we can leverage in this determination are: (a) who is posting the post, and (b) what is the context of the post.

    %\\*\textbf{Classifier}
    %Hereto, we elaborate more on the details of the classifier by explaining each section seperately.
    {\bf Step 1: Behavioral Features.} We associate each user of the forum with a set of 11 features that capture their behavior. In particular: 
    \begin{itemize}
        \item Number of posts; the total number of posts made by the user
        \item Number of threads; the total number of threads the user has contributed to
        \item Number of threads initiated; the total number of threads initiated by the user
        \item Average thread entropy; the average entropy of the user distribution of the threads in which the user has contributed to
        \item Number of active days; the number of days that the user generates at least one post
        \item Average day entropy; the average entropy of the user distribution of the posts made on the days that the user is active
        \item Active lifetime; the number of days between the first and the last post of the user
        \item Wait time; the number of days passed between the day the user joined the forum and the day the user contributed their first post
        \item Average post length; the average number of characters in the user's posts
        \item Median post length; the median number of characters in the user's posts
        \item Maximum post length; the number of character's in the user's longest post
    \end{itemize}

    {\bf Step 2: Contextual Features. }
    Apart from the aforementioned behavioral features we also include features related with the context in which an IP address appears within a post. 
    In particular, we consider the frequency of the words (except stop-words) in the posts. 
    Words that are frequent only in few documents (posts in our case) are more informative than those that appear frequently on a larger corpus \cite{Ramos2003UsingTT}.  
    To this end, we use TF-IDF to weight the various words/terms that appear in our data.  
    %\mrem{M: Is this the only thing we are using? How many features do we use? Can we describe at least roughly?}\reminder{done}
    %
    %\mrem{I do not understand what the following says ---
    %We have weight of more than \textit{10k} words then selecting informative keywords %and extracting latent feature sets are done in the next step as following.
    %}
    %
    After calculating the frequency and the corresponding weights of each word in the dataset we end up with more than 10,000 features/terms. 
    Hence, in the next step we select discriminative features by extracting latent features.

    %\item \textbf{\textit{Data Labeling}}: We assign a \textit{Malicious} label to a post containing an IP as  if the link was reported in Virustotal. Otherwise, we label them as \textit{Benign}. 
    %\item \textbf{\textit{Balanced Dataset}}: 

    %In order to utilize this model as a warning system to give heads-up about the potential malicious IP addresses we have selected all post containing IP which date of appearance in the forum is before reported on Virustoatal. Having the labels of the data, we create a training dataset based on equal number of \textit{Malicious} and \textit{Benign} posts. The number of final post in the train dataset varies in different dataset based on how many malicious IP will be find in each by \textit{VirusTotal} API. 
    %\item \textbf{\textit{Feature selection and extraction}}:
    %\reminder{Vagelis: please check}
    %\reminder{From Vagelis: should we say explicitly that our feature selection involves two methods, infoGain and co-clustering? Maye name this paragraph ``feature selection''?}
    %Extracting all the features mentioned in \textit{features} section ends up with tons of words as features in addition to behavioral features. 
    We begin by performing feature selection in order to identify the most informative features by applying the information gain framework \cite{Yang:1997:CSF:645526.657137}.  
    %For a classifier some of these features might not be informative in the model so it is worth to decrease the number of undiscriminating features. To this end, we apply information gain as a well-known method for features selection\cite{????}. 
    Furthermore, in order to avoid overfitting we pick a random subset of posts from the whole dataset and select the highest ranked features based on \textit{Information Gain} score. In this way, a subset of discriminative keywords, 100 in our model, are selected. % ,i.e. 100 keywords in our model. 
    %\reminder{Vagelis please check this can be implied a potential novelty of this model} 
    It turns out that each user uses only a small number of those words, resulting in a sparse dataset which we wish to exploit in our model. 

    \begin{table*}
    \normalsize
    \renewcommand{\arraystretch}{1.3}
    \centering
    \caption{Selecting a classifier: overall accuracy.}
         
	\begin{tabular}{c|c|c|c}
    \hline
    Forum & Naive Bayes & 3NN & Logistic regression \\
    \hline
    \WS & 91.9\% & 87.1 \% & 94.8\%  \\
    \OffCommShort & 84.1\% & 83.2\% & 86.5\%  \\
    Ashiyane & 85.1\% & 82.3\% & 94\%  \\
    \hline
	\end{tabular}
	\label{tab:compareClassifiers}
	\end{table*}

    \begin{table*}
 	\normalsize
    \renewcommand{\arraystretch}{1.3}
    \centering
    \caption{\myalg evaluation: 10-fold cross validation evaluation (using Logistic Regression).}
    \label{tab:Train_result}      
	\begin{tabular}{c|c|c|c|c}
    \hline
    Forum & Instances  & Precision & Recall  & ROC Area \\
    \hline
    \WS &       362 & 0.9   & 0.94 & 0.96 \\
    \OffCommShort & 342  & 0.88  & 0.85 & 0.91 \\
    \Ash &     446  & 0.9   & 0.92 & 0.92 \\
    \hline
	\end{tabular}
	\label{tab:bayes}
	\end{table*}

    {\bf Step 3: Identifying \blfsets.}  %\mrem{@V: Can you fix this?}
    %\reminder{@M: I simplified it a bit}
    We also like to leverage latent similarities of different posts in some of the dimensions spanned by post features and behavioral features for the writer of the post. Essentially, we seek to identify groups of highly similar posts under a small number of features, which does not necessarily span the full set of features. The reason why we wish to pinpoint a subset of the features instead of the entire set is because this way we are able to detect subtle patterns that may go undetected if we require post similarity across all the features. We call those sets of feastures \lfsets. To this end, we apply a soft co-clustering method, Sparse Matrix Regression (SMR) \cite{papalexakis2013k}, to exploit the sparsity and extract latent features of the post containing IP addresses. Given a matrix $\mathbf{X}$ of posts $\times$ features, its soft co-clustering via SMR can be posed as the following optimization problem:

 \begin{math}
        \min_{\mathbf{a}_r\geq0,\mathbf{b}_r\geq0} \| \mathbf{X} - \sum_r^R \mathbf{a}_r \mathbf{b}_r^T\|_F^2  + \lambda \sum_{i,r}|\mathbf{a}_r(i) | +\\ \lambda \sum_{j,r}|\mathbf{b}_r(j) |
\end{math}
    
    %where $\mathbf{a}_r$ is the post to $r$-th co-cluster assignment, and $\mathbf{b}_r$ is the feature to $r$-th co-cluster assignment. 
    where  $\mathbf{a}_r$ and  $\mathbf{b}_r$ are vectors that ``describe'' co-cluster $r$, which we explain below.
    %The way to interpret the results of SMR is the following: 
    Each $\mathbf{a}_r$ is a vector with as many dimensions as posts. Each value $\mathbf{a}_r(i)$ expresses whether post $i$ is affiliated with co-cluster $r$.
    Similarly,  $\mathbf{b}_r$ is a vector with as many dimensions as features, and $\mathbf{b}_r(j)$ expresses whether feature $j$ is affiliated with with co-cluster $r$.
        %Because we impose sparsity through  $\ell_1$ norm regularization, we expect the majority if those coefficients to be exactly zero, indicating that a particular post does not belong to a co-cluster.
    Parameter $\lambda$ controls how sparse the co-cluster assignments are, effectively controlling the co-cluster size. As we increase $\lambda$ we get sparser results, hence cleaner co-clustering assignments. We tune $\lambda$ via trial-and-error so that we obtain clean but non-empty co-clusters, and we select $\lambda = 0.01$ in our case.
    %\mrem{What value did we select here?}\reminder{done}
    %Intuitively, $\lambda$ has to be large enough for all of the assignments to be non-empty but also relatively noise-free. 
    
   % \reminder{Vagelis please help: maybe more distribution needed to justify how \mathbf{a} matrix shows latent features}
    
    %\item \textbf{\textit{Traning the model}}:
{\bf Step 4: Training the model.} We subsequently train a number of classifiers using the selected features based on $\mathbf{a}$ matrix. 
In particular, we examine (a) a Naive Bayes classifier, (b) a K-Nearest Neighbor classifier  and (c) a logistic regression classifier. 
Our 10-fold cross validation indicates that the Logistic regression classifier outperforms kNN and Naive Bayse, achieving high accuracy, precision and recall (see Table \ref{tab:compareClassifiers}).

    {\bf Determining feature sets.}
    We investigate the effect of selecting different feature sets in classifying IP addresses in forums. To this end, we investigate three subsets of the features  discussed earlier.
    
    {\bf a. \WordFreq} is the normalized frequency of the most informative words that appear in a post as discussed in Step 2.
    
    {\bf b. \Mixed} is the set of features which consists of the combination of the words frequency features, defined above, and user behaviour features, which are extracted in Step 1. In other words, it is the union of the features in Step 1 and Step 2.
    
    {\bf c. \CoCluster} is the latent set of features extracted in Step 3 by applying the co-clustering approach on the \Mixed features set.
    
\begin{figure}[h]
    \includegraphics[width=\linewidth]{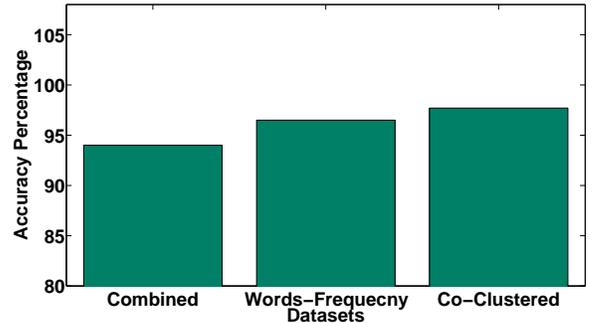}
    \caption{Accuracy of different feature sets in \WS forum to detect malicious IP}
    \label{fig:malIPFeatures}
    \end{figure}

    We evaluate these three sets of features on their ability to enable the classification. In more detail,
    we use these features with a classifier to assess their effectiveness by computing the accuracy of the classifier to identify malicious IP addresses. According to the results which are shown in Figure \ref{fig:malIPFeatures}, the \CoCluster features set exhibits higher accuracy by 4.1\%  compared to \WordFreq. On the other hand, although the \Mixed features do not increase the accuracy compared to the \WordFreq,  the co-clustering method  does. It extracts the latent features from the \Mixed features set and outperforms \WordFreq and \Mixed in identifying malicious IP addresses.
    % }
    
    \subsection{Applying \myalg on the forums}
    
    Having established the statistical confidence of our classifier, we  apply it on the posts of the forums except the ones that we used in our groundtruth.
    We use the logistic regression classifier as it exhibits the best performance.
    
    % {\color{red} 
    Applying \myalg on the forums shows that there is a wealth of information
that we can extract from security forums in two aspects of the quantity and time of detecting malicious IP against VirusTotal.

    {\bf a. Detecting more IP addresses.}
        With \myalg, we find an additional  670 malicious IP addresses  in \WS, and 617    in \OffComm  806 in \Ash (see Table \ref{tab:test_result}).  
    In other words, \myalg enables us to find three times additional malicious IP addresses in total compared to the IP addresses found on VirusTotal. 
    It is interesting to observe that this factor varies among our three sites.
    For \Ash, our method finds roughly 6 times additional malicious IP addresses.
    With a precision of roughly around 90\% and considering small amount of \textit{False Positive} rate, our method can add a significant
    number of malicious IP addresses to a blacklist. Using the limited manual inspection, we confirm that the precision of the method on out of sample data is in the order of 88\%.
    
    {\bf b. Detecting malicious IP addresses earlier: more than half IPs, at least 3 months earlier}
    Here we focus on the malicious IP addresses that are jointly identified by our method and VirusTotal and compare the time that they were reported in each source,
    and show the results in  Table \ref{tab:EarlyDetection}
    for 3, 6 and 12 months difference in time.
         We compare jointly detected IP addresses with \myalg and VirusTotal in terms of time that the IP addresses were mentioned in posts and the time they were reported on VirusTotal. 
    We see that on average 62\% of the malicious IP addresses with \myalg could be identified at least 3 months earlier than VirusTotal. 
    We can see that with \myalg,  we find 53\%, 71\% and 62\% of these IP addresses   in \WS, \OffComm and \Ash  respectively at least 3 months earlier than  in VirusTotal.
    We also identify 39\% and 24\%  of the malicious IP addresses respectively at least 6 and 12 months earlier with \myalg.
    
    {\bf Additional stress-testing of our accuracy:} In order to assess the performance of our approach,  we randomly picked 10 percent of the labeled data with \myalg method and annotated them manually by human annotators. The calculated accuracy on the sampled data shows more than 85\% accuracy on average over all datasets which is close but somewhat lower than the reported accuracy in the Table \ref{tab:compareClassifiers}.

    \begin{figure*}[ht]
\begin{minipage}{0.32\linewidth}
  \includegraphics[width=\linewidth]{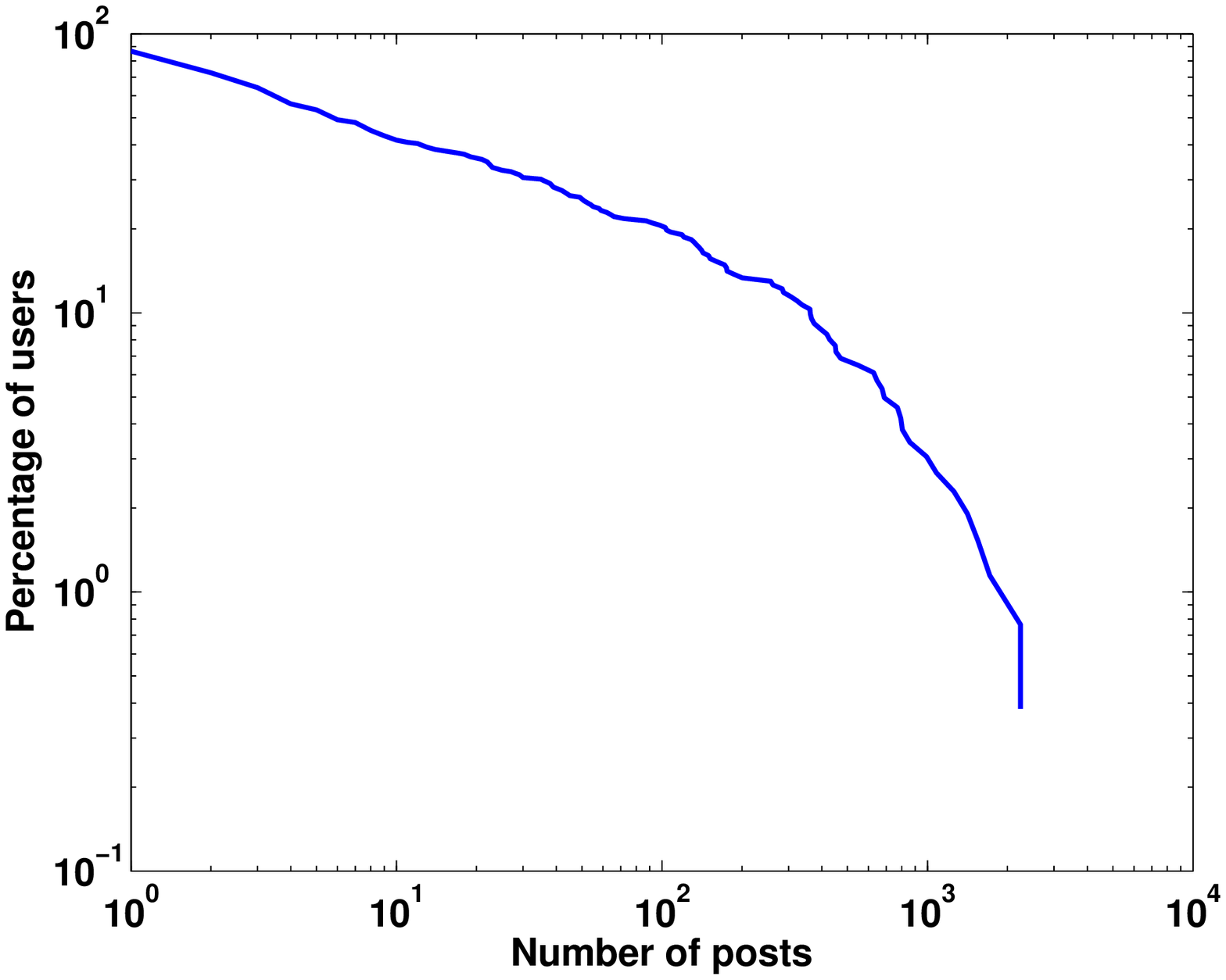}
  \centering
  {(a) \WS}\label{fig:MalIPUsertWS}
\end{minipage}
\begin{minipage}{0.32\linewidth}
  \includegraphics[width=\linewidth]{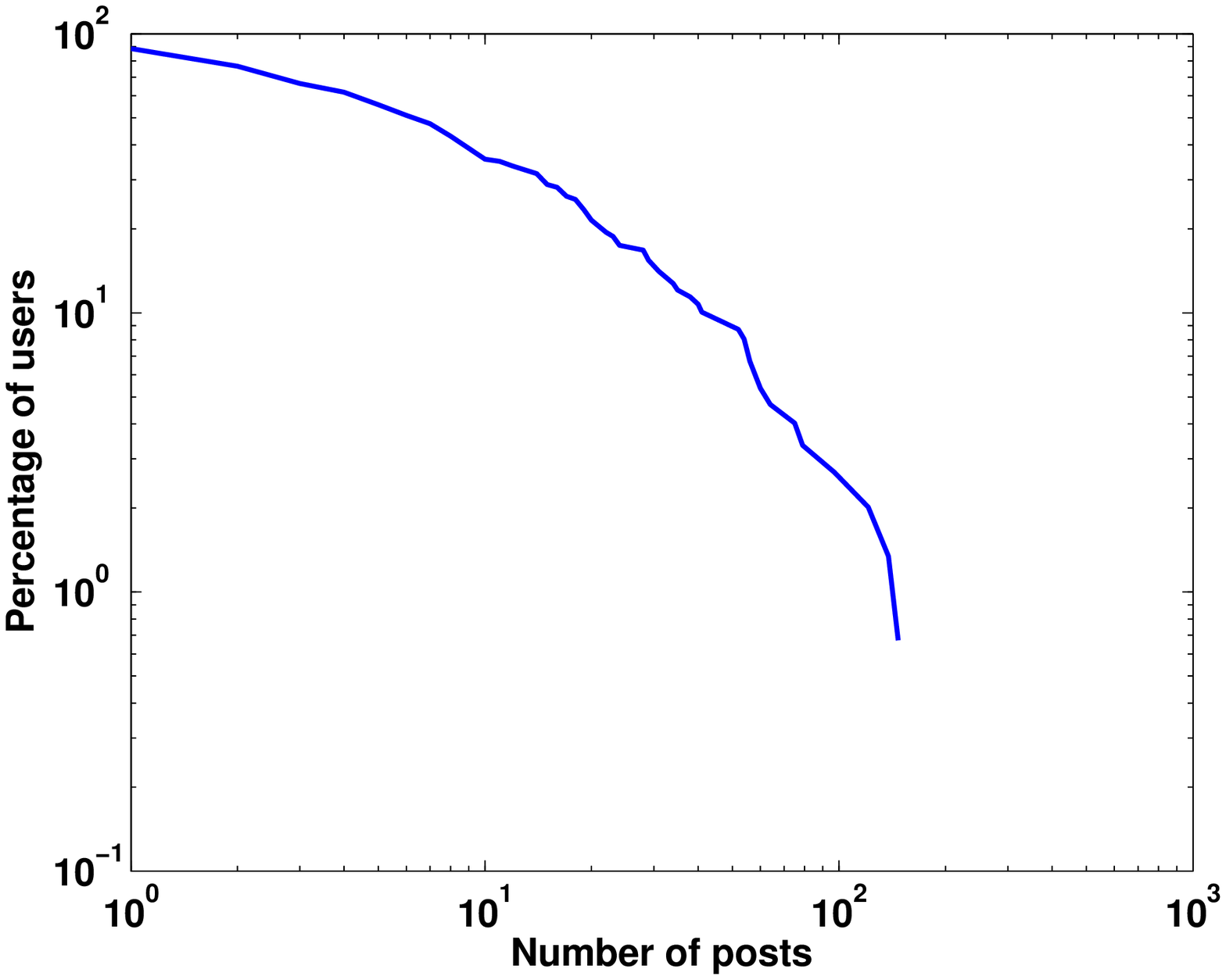}
  \centering
  {(b) \OffComm}\label{fig:MalIPUserOffComm}
\end{minipage}
\begin{minipage}{0.32\linewidth}%
  \includegraphics[width=\linewidth]{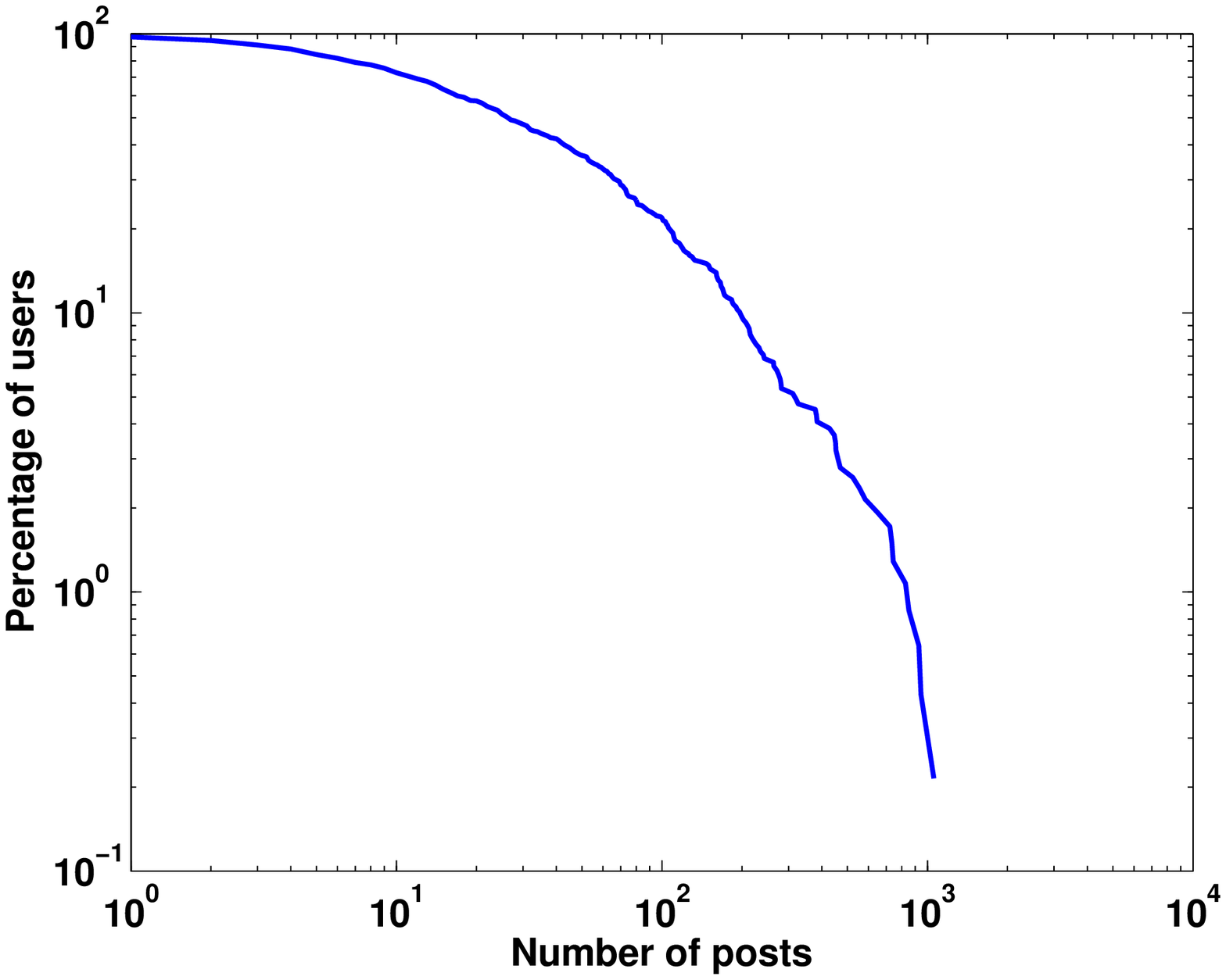}
  \centering
  {(c) \Ash }\label{fig:MalIPUserAsh}
\end{minipage}
\caption{CCDF of the number of overall posts per \Malicious users (who report malicious IPs)  in $\log$-$\log$ scale.}\label{fig:MalIPUser}
\end{figure*}

%    \begin{figure}[h]
%    \includegraphics[width=\linewidth]{figures/CompareTime.png}
%    \caption{Distribution of the early detecting malicious in \myalg against VirusTotal}
%    \label{fig:compareTime}
%    \end{figure}   

\iffalse
\begin{figure*}[!htb]
\begin{minipage}{begin}{0.32\linewidth}
  \includegraphics[width=\linewidth]{earlydetection_cdf_wilders.eps}
  \centering
  {(a) \WS }\label{fig:EarlyWS}
\end{minipage}
\begin{minipage}{begin}{0.32\linewidth}
  \includegraphics[width=\linewidth]{earlydetection_cdf_offcomm.eps}
  \centering
  {(b) \OffComm}\label{fig:EarlyOffComm}
\end{minipage}
\begin{minipage}{begin}{0.32\linewidth}%
  \includegraphics[width=\linewidth]{earlydetection_cdf_ashiyane.eps}
  \centering
  {(c) \Ash}\label{fig:EarlyAsh}
\end{minipage}
\caption{CCDF of the number of malicious IP detected in \myalg earlier than VirusTotal ($\log$-$\log$ scale).}\label{fig:EarlyDetection}

\end{figure*}
\fi

    \begin{table}[ht]
   	\normalsize
    \renewcommand{\arraystretch}{1.3}
    \centering
    \caption{Timely comparison between jointly detected malicious IP addresses in \myalg and VirusTotal. Reported percentage of malicious IP Addresses which \myalg detected earlier than VirusTotal }
    \label{tab:EarlyDetection}      
	\begin{tabular}{|c|c|c|c|}
	\hline
    \multicolumn{1}{|c}{} & \multicolumn{3}{|c|}{ At least X months earlier} \\
    \hline
    Dataset & 3 &  6 & 12 \\
    \hline
    \WS & 53\%   & 23\% & 14\% \\
    \OffComm & 71\%  & 46\% & 21\%\\
    \Ash & 62\%  & 49\% & 37\% \\
    \hline
    Average (across forums) & 62\%  & 39\% & 24\% \\
    \hline
	\end{tabular}
\end{table}

{\bf \Malicious Users.}
    Who are the users that report malicious IP addresses? We want to understand and ideally, develop a profile for these users, which
    we will refer to as {\bf \Malicious} users. We start by considering the number of post these users  post on the forums.
    
    {\bf The majority of IP reporting is done by highly active (more than 10 posts overall) in \Ash}. In Figure \ref{fig:MalIPUser}, we show the cumulative complementary distribution function for the number of posts per \Malicious user for \Ash. More than 72\% of the \Malicious users post more than 10 posts overall, which we consider as high engagement given the distribution of posting that we saw in the previous section. Therefore, in \Ash, \Malicious users are contributing significantly in reporting malicious IP addresses. Intrigued, we examined further and found that, among them, there are two users who have more than 1000 posts,
    1058 and 2780 to be exact, and whose user-names are \textit{"Classic"} and \textit{"Crisis"}. 
    On the other side of the spectrum, 2.4\% of \Malicious users  have posted a single post in the forum, and in that post they reported a malicious IP address.
    
    {\bf The majority of IP reporting is done by less active users (less than 10 posts overall) in \OffComm}.
        In Figure \ref{fig:MalIPUser}, we show the cumulative complementary distribution function for the number of posts per \Malicious user for  \OffComm.  Unlike \Ash, here 65\% of the \Malicious users have less than 10 posts overall. 
        % On the other hand, \textit{regular} users, who are the users with moderate number of posting(i.e. (2-10) posts) on the forum, are posting most of the malicious IP addresses. In \OffComm 53\% of the \Malicious users are in this category.
        Going into more detail, roughly 12\% of the \Malicious users have a single post overall, while 26\% of them have only two overall posts. 
        The same behavior is observed in  \WS which is shown in Figure \ref{fig:MalIPUser}. 
        
        Overall, there does not seem to be an obvious
        pattern between number of total posts and number of malicious IPs reported among \Malicious users.
    
    \subsection{Case-study: from reported malicious IPs to a DDoS attack}
    
    We show that mining the forums could actually provide information about real events.  We identify a link between a malicious IP address that
    our method detected with an actual DDoS attack.
    
\begin{figure}[htb!]
\centering
\includegraphics[width=\linewidth]{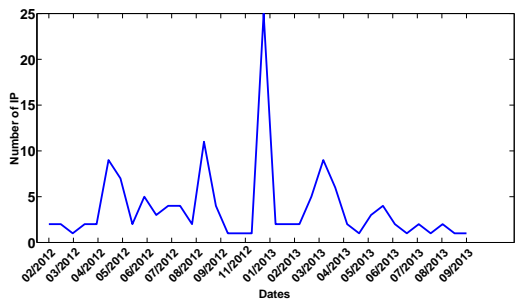}
\caption{Time-series of the number of posts containing malicious IP reported in each month for \WS.}
\label{fig:MalIPTimeSeries}
\end{figure}

    We conducted the following analysis.
    We plot the time-series of the number of posts containing malicious IP addresses in \WS  from 2012 to 2013 found by \myalg. We show the  time-series in Figure \ref{fig:MalIPTimeSeries}. We observe some spikes on these time-series, which we further analyze. One of the spikes 
    was in September 2012, and it reports a set of malicious IP addresses that were involved in an DDoS attack that month. That same thread continued being active, and in December of 2012, it was reported in that thread that attack was caused by \textit{Nitol Botnet} due
     to a Microsoft's vulnerability~\cite{nitol}.
    
    We argue that this case-study points to additional layers of functionality that can be built upon our method, that can provide
    a semi-automated way to extract richer information beyond just reporting  malicious IP addresses.
    
%}

 \subsection{ Discussion and limitations}
 
  Although our method  exhibits pretty good accuracy overall, 
  we attempt to understand its limitations and  detect
  the source of misclassifications. 
 
 {\bf Limited text in the post:} 
     The words in the post provide significant evidence for the classification. In some cases, some posts are very sparse in their text, which makes the classification of the included IP address harder.
     We consider these kinds of posts a significant contributor to misclassifications.
 
 {\bf Characterization at the post level:}
 In our method, we classify an IP address by using features at
 the level of a post. 
 Recall that roughly 86\% of all posts across all forums has a single IP per post as shown in Figure \ref{fig:IPPerPost}.
  In other words, having more than one IP address per post is already not very common.
  Furthermore, even more rarely, we have seen a few cases, where a post contains both a benign and a malicious IP address.
  %\reminder{add example}
 As our method is currently set-up, this will lead to errors in the classification. A straightforward solution is to  consider examining 
 the text surrounding each IP address within the post.

\section{  SpatioTemporal Analysis}
\label{sec:spTemp}

 In this section, we discuss the spatiotemporal features of the  malicious IP addresses identified in security forums
 in Section \ref{sec:ipmodel}.

 \subsection{Temporal analysis}
   
   The key question from a temporal point of view is if the number of reported
   malicious IP addresses increases or decreases over time.

     \begin{figure}[h]
    \includegraphics[width=\linewidth]{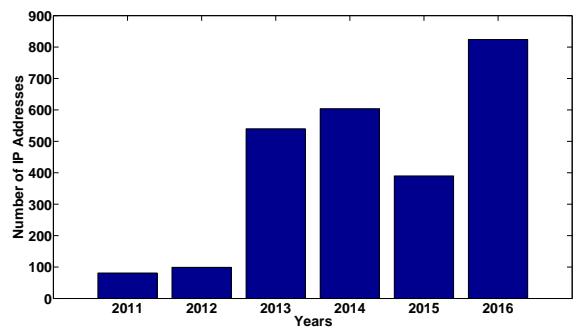}
    \caption{Increasing trend: Malicious IP addresses reported on the forums  each year.}
    \label{fig:IPTime}
    \end{figure}
    
    {\bf The number of reported malicious IP addresses has
increased  by  a  factor  8  in  4  four  years.}
In Figure \ref{fig:IPTime}, we plot the number of reported malicious IPs found by our method across all three forums between 2011-2016. We find that the number increased by a factor of 8: from roughly 100 to roughly 800.  In spite of some decreases in years 2011, 2012 and 2015, it has a clear increasing trend.

\subsection{Spatial analysis}
 We study the geo-location of the identified IP addresses
 from Section \ref{sec:ipmodel}. 
%  We  consider the geographical and time distribution of malicious IP addresses and compare it in different forums.
We utilize \textit{GeoLite} database \cite{GeoLite}, which can show us
the country and continent of an  IP address.
Here we focus on continents of the IP addresses location.
% According to the distributions depicted in the Fig \ref{fig:LocTime} and Figure \ref{fig:IPTime} we make the following observations. 

A natural question to ask is whether the geographical distribution of the malicious addresses differs between VirusTotal and \myalg. We investigate this in detail below.

\begin{figure*}[!htb]

\begin{minipage}{0.49\linewidth}
  \includegraphics[width=\linewidth]{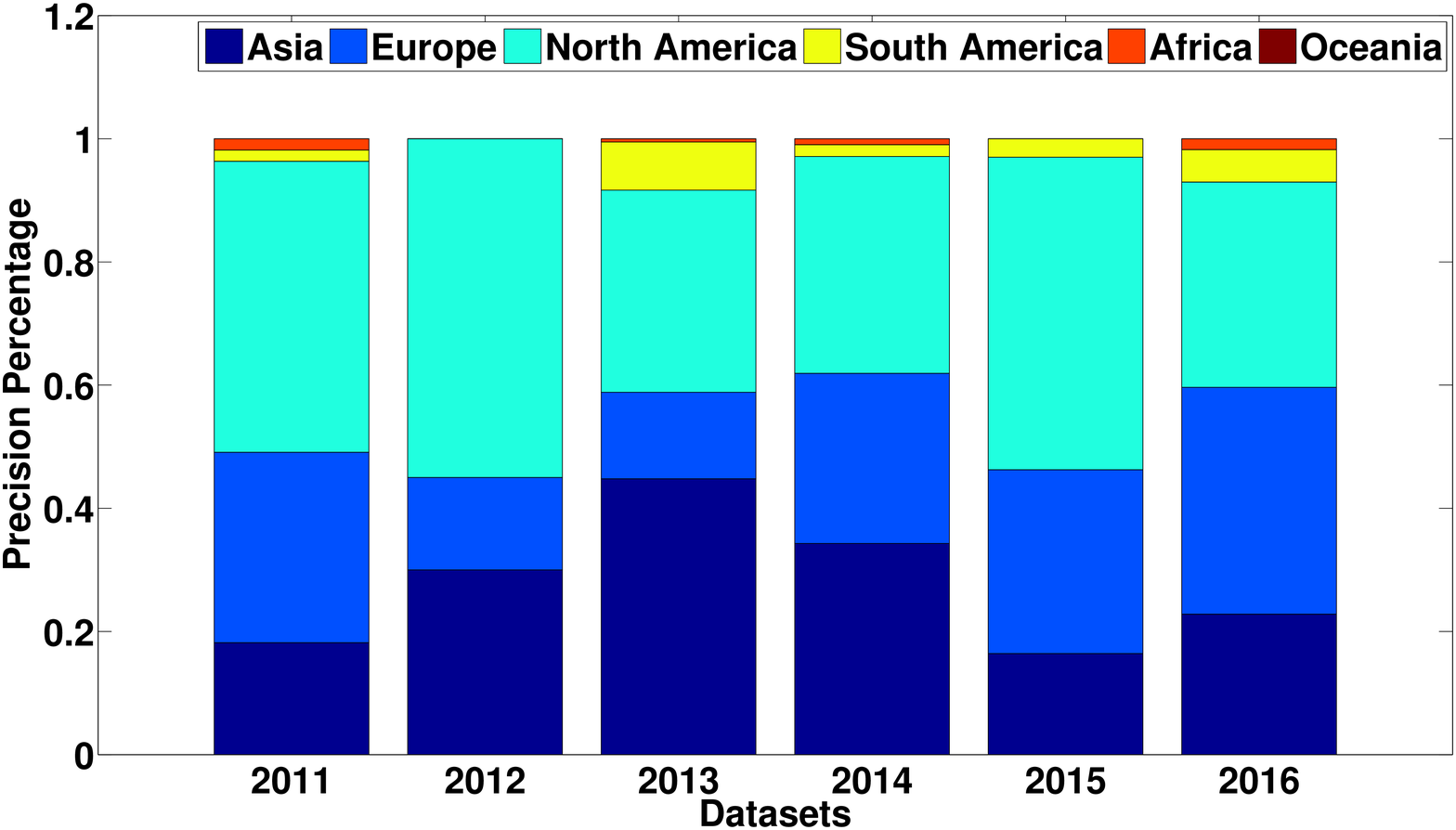}
  \centering
  {(a) VirusTotal }\label{fig:LocTimeVT}
\end{minipage}
%\minipage{0.33\linewidth}
%  \includegraphics[width=\linewidth]{figures/LocTimeInferIP.eps}
%  \centering
%  \subcaptionbox{(b) \myalg}\label{fig:LocTimeInferIP}
%\endminipage
\begin{minipage}{0.49\linewidth}
  \includegraphics[width=\linewidth]{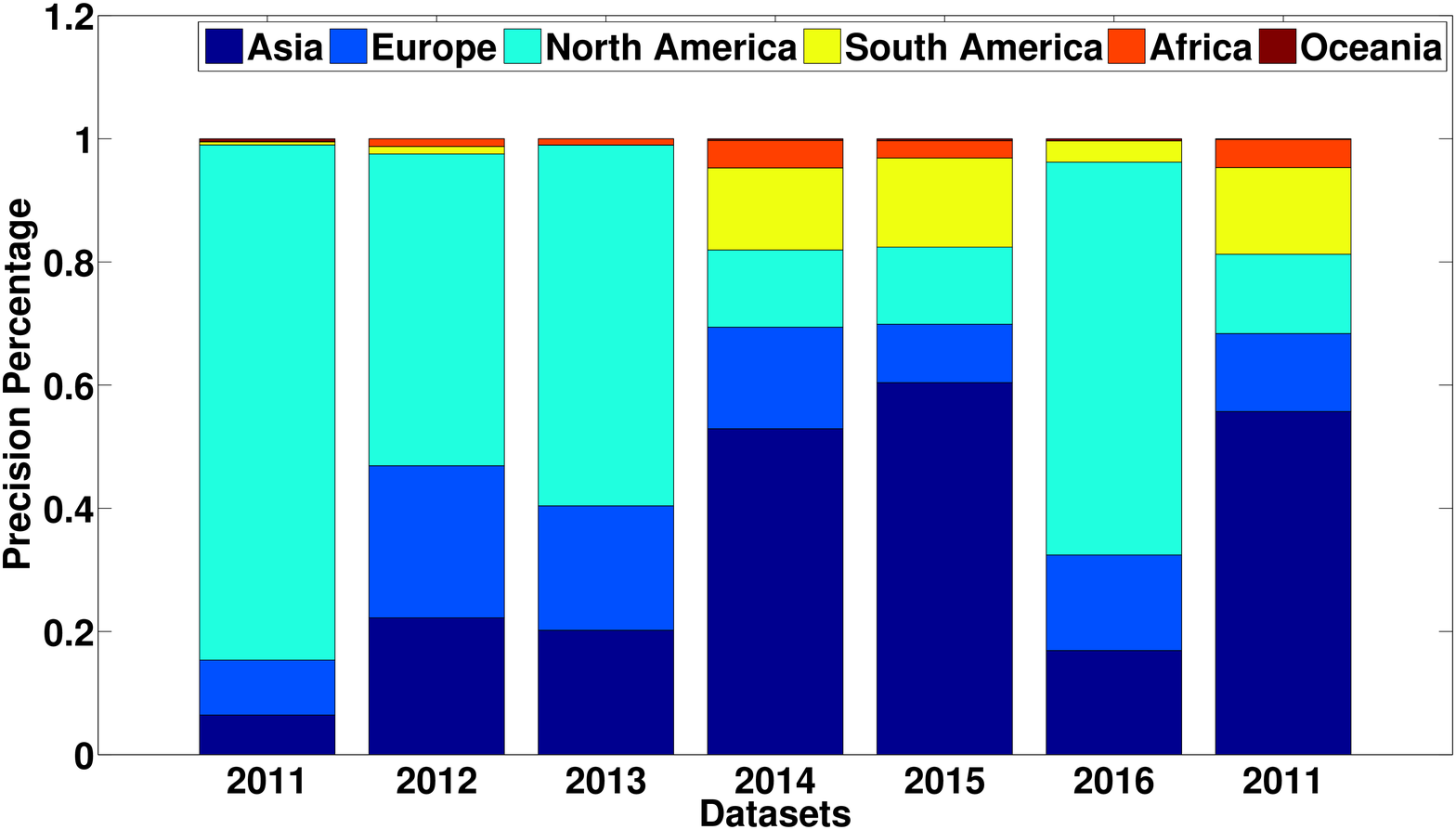}
  \centering
  {(b)\myalg }\label{fig:LocTimeV}
\end{minipage}
\caption{SpatioTemporal distribution of malicious IP addresses detected by \myalg and VT .}\label{fig:LocTime}
\end{figure*}

{\bf  VirusTotal:   North America hosts the majority of the reported malicious IP addresses.} We plot the percentage of the distribution of the IP addresses extracted from VirusTotal across continents in Figure \ref{fig:LocTime} (a) between 2011-2016. We observe that the majority of the malicious IP addresses are located in the North America continent. There are two exception in 2013 and 2016 when  Asia and Europe respectively contain most of the malicious IP addresses. Overall, Table \ref{tab:AvgDist} shows
the geo-graphical distribution over all the years: North America, Asia and Europe are the three most active continents in that order.
    
\begin{figure}[htb]
 \centering
\includegraphics[width=\linewidth]{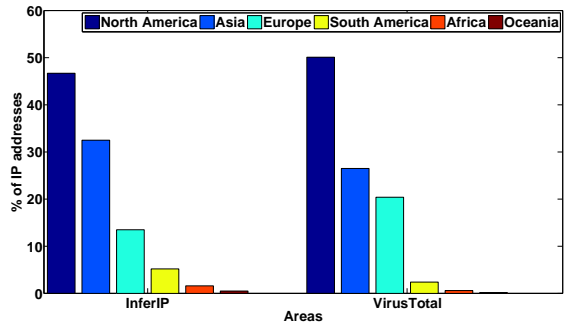}
\caption{The percentage distribution of malicious IP Addresses in each continent across all three forums for \myalg and VirusTotal.}
\label{fig:MalIPLoc}
\end{figure}

{\bf \myalg:  North America dominates again, but South America and Africa have non-trivial contributions.}
We plot the percentage of the distribution of the IP addresses extracted form \myalg across continents in Figure \ref{fig:LocTime} (b) between 2011-2016. We observe that North America hosts the majority of the reported malicious IP addresses again, but
we find a more diverse global activity compared to what we observed in VirusTotal. For example, we can see that in years 2013, 2014, and 2016: (a) Asia has the majority of the malicious IP addresses, and (b) South America and Africa have a considerable percentage of malicious IP addresses. However,  when seen across all years, the geographical distributions of the IPs in \myalg and VirusTotal quite similar:  North America, Asia and Europe have the majority of the malicious IPs detected by \myalg  similarly to those of VirusTotal. In Figure \ref{fig:MalIPLoc}, 
we plot the geographical distribution of malicious IPs per continent across 
all years and all forums for \myalg and VirusTotal,
while the exact numbers are shown in Table \ref{tab:AvgDist}.
Qualitatively the distributions look relatively similar, especially in the order of significance of the continents, but at the same, we can see that South America and Africa have a larger percentage of IP addresses in \myalg compared to those in VirusTotal.

\begin{table*}
\normalsize
 \centering 
 \caption {Percentage of distribution of IP addresses across continents over all the years.}
\label{tab:AvgDist}
 \begin{tabular}{| c | c | c | c | c| c | c|} 
 \hline
  & North America & Asia & Europe  & South America & Africa & Oceania\\
 \hline
 \myalg & \textbf{46.7} & 32.5 &	13.5 & 5.2 &	1.6 & 	0.5
 \\ 
 VirusTotal & \textbf{50} & 26.5 & 20.4  & 2.4 & 0.6 & 0.17 \\
 \hline
\end{tabular}
\end{table*}

% \begin{figure}[h]
%   {\includegraphics[width=\linewidth]{figures/WildersTS.png}}
% \caption{Malicious IP Addresses found in \WS from 2012 to 2017}
% \label{fig:WilderTS}
% \end{figure}

% \begin{figure}[h]
%   {\includegraphics[width=\linewidth]{figures/OffCommTS.png}}
% \caption{Malicious IP Addresses found in \OffComm from 2012 to 2017}
% \label{fig:OffCommTS}
% \end{figure}

    %  Since the extracted features discussed in Section \ref{sec:ipmodel} (\WordFreq, \Mixed and \CoCluster), related to IP addresses are in the post level mean we extract features for the post containing the IP addresses. Therefore if more than one IP address exist in a post we must label them identically.

\section{Related Work}
\label{sec:related}

%\subsection{Related Studies}
  
  We briefly discuss three categories of relevant research. 
  
   {\bf a. Analyzing structured security sources.} There is a long line of research studying the ontology of cyber security and the automatic extraction of information from structured security documents. 
    Iannacone {\em et al.}\cite{Iannacone2015} developed a schema for extracting relevant concepts from various types of structured data sources. In another work, Blanco {\em et al.} \cite{Blanco2008} proposed methods to detect anomalies on the extracted ontology and network flow graph. Moreover,  Bridges {\em et al.}\cite{Bridges2013} proposed a method to do entity labeling on structured data by utilizing neural networks.
 These work are complementary to ours as we focus on unstructured data,
 which poses different challenges. 
 
 {\bf b. Analyzing online security forums.} Recently security forums have been the focus of various studies that showcase the usefulness of the information present in security forums. For example, Motoyama {\em et al.} \cite{Motoyama:2011} present a comprehensive statistical analysis in underground forums. 
 Others studies focus on the users’ classification  or the discovery of the relationships between the forum’s members \cite{Zhang:2015,Abbasi2014}. 
Extracting different discussion topic in the forums and classifying the language of the codes posted in the forum has been done in \cite{Az1}. 
 Contrary to these studies, our work emphasizes on the development of automated systems that actually exploit the wealth of information in these security forums in order to enhance security. Similar to detecting malicious users on commenting platforms has been done on \cite{mike2017}. 
  A recent work analyzes security forums  to identify and geo-locate Canadian IP addresses focusing on spam
 and phishing \cite{Frank2016} and in another work, Portnoff {\em et al.} \cite{Portnoff2017} studies the exchange of  malicious
 services and tools and studies their prices on the security forums.

 %/iffalse
 {\bf c. Analyzing blogs and social networks.} There has been a plethora of studies on blogs and social media, but their goals are typical not related to extracting security information.
 \cite{leskovec2017,Althoff:2017,Johan2011}. The studies range from modeling user behavior \cite{Devineni:2015:WTP:2808797.2808880,Reza2016} to inferring information about the user (demographics, preferences, mental state), and to modeling the information propagation on online forums.
 Although interesting, the focus of these studies are significantly different from our goal here.
 %\fi
 
\section{Conclusion}
\label{sec:conclusions}

The  take away message from our work is that there seems to be a wealth of useful information in security forums. The challenge is that the information
is unstructured and we need novel methods to extract it.
In this direction, a key insight of our work is that using behavioral and text-based features
can provide promising results.

In support of this assertion, we develop a systematic method to extract malicious IP addresses  reported in security forums.  
We utilize both behavioral, as well as textual features and show that we can detect malicious IP addresses with high accuracy, precision and recall.  
Our results in Table \ref{tab:test_result} are promising.

We then apply \myalg to all the posts we have collected.
Although are classification is not perfect,
our relatively high precision (hovering around 90\% in Table \ref{tab:Train_result}) provides sufficient confidence in our results.
We find three times as many additional malicious IP addresses as the original malicious IP addresses identified by VirusTotal.
Furthermore, even for the jointly discovered IP addresses,  at least 53\% of the IP addresses detected at least 3 months earlier than VirusTotal. 
The key message from our spatiotemporal analysis is that the number of reported malicious IP addresses is increasing over time. 
 
%Overall, this is strong evidence for the ability of security forums to enhance the response time of traditional blacklist services.  

In the future, we plan to extend our work by extracting other types of security information. 
Our first goal is to   detect malicious URLs mentioned in the forums.
Our second and more ambitious goal is to identify the emergence of new malware, threats, and possibly attacks, which we expect to see associated with large numbers of panic-filled or help-requesting posts.
Our final goal is to identify malicious users, since interestingly,
some users seem to be promoting and  selling hacking tools in these forums.

%We will further explore this direction in our future work.

\section{Acknowledgments}
\label{sec:Acknowledge}

%\thanks {This work was supported by  DHS ST Cyber Security (DDoSD)  HSHQDC-14-R-B00017 grant and the Bourns College of Engineering at UC Riverside}
%Any opinions, findings, and conclusions or recommendations expressed in this material are those of the author(s) and do not necessarily reflect the views of the funding parties.

\thanks {%The authors thank the anonymous reviewers for their useful comments.
This material is based upon work supported by an Adobe Data Science Research Faculty Award, and DHS ST Cyber Security (DDoSD)  HSHQDC-14-R-B00017 grant. Any opinions, findings, and conclusions or recommendations expressed in this material are those of the author(s) and do not necessarily reflect the views of 
%the National Science Foundation or other 
the funding institutions.}

%\vspace{-0.1in}
%\small
\bibliographystyle{abbrv}
%\vspace{0.2in}
\bibliography{ref}

\begin{thebibliography}{10}

\bibitem{ashiyane}
Ashiyane.
\newblock \url{http://www.ashiyane.org/forums/}.

\bibitem{GeoLite}
Geolite.
\newblock \url{http://dev.maxmind.com/geoip/legacy/geolite/}.

\bibitem{nitol}
Nitol-botnet.
\newblock \url{https://threatpost.com/tag/nitol-botnet/}.

\bibitem{offcomm}
Offensive community.
\newblock \url{http://www.offensivecommunity.net}.

\bibitem{virustotal}
Virustotal.
\newblock \url{http://www.virustotal.com}.

\bibitem{wilders}
Wilders security.
\newblock \url{http://www.wilderssecurity.com}.

\bibitem{Abbasi2014}
A.~Abbasi, W.~Li, V.~Benjamin, S.~Hu, and H.~Chen.
\newblock Descriptive analytics: Examining expert hackers in web forums.
\newblock In {\em 2014 IEEE Joint Intelligence and Security Informatics
  Conference}, pages 56--63, Sept 2014.

\bibitem{Althoff:2017}
T.~Althoff, P.~Jindal, and J.~Leskovec.
\newblock Online actions with offline impact: How online social networks
  influence online and offline user behavior.
\newblock In {\em Proceedings of the Tenth ACM International Conference on Web
  Search and Data Mining}, WSDM '17, pages 537--546, New York, NY, USA, 2017.
  ACM.

\bibitem{Blanco2008}
C.~Blanco, J.~Lasheras, R.~Valencia-García, E.~Fernández-Medina, A.~Toval,
  and M.~Piattini.
\newblock A systematic review and comparison of security ontologies.
\newblock In {\em 2008 Third International Conference on Availability,
  Reliability and Security}, pages 813--820, March 2008.

\bibitem{Bridges2013}
R.~A. Bridges, C.~L. Jones, M.~D. Iannacone, and J.~R. Goodall.
\newblock Automatic labeling for entity extraction in cyber security.
\newblock {\em CoRR}, abs/1308.4941, 2013.

\bibitem{leskovec2017}
J.~{Cheng}, M.~{Bernstein}, C.~{Danescu-Niculescu-Mizil}, and J.~{Leskovec}.
\newblock {Anyone Can Become a Troll: Causes of Trolling Behavior in Online
  Discussions}.
\newblock {\em ArXiv e-prints}, Feb. 2017.

\bibitem{Devineni:2015:WTP:2808797.2808880}
P.~Devineni, D.~Koutra, M.~Faloutsos, and C.~Faloutsos.
\newblock If walls could talk: Patterns and anomalies in facebook wallposts.
\newblock In {\em Proceedings of the 2015 IEEE/ACM International Conference on
  Advances in Social Networks Analysis and Mining 2015}, ASONAM '15, pages
  367--374, New York, NY, USA, 2015. ACM.

\bibitem{Frank2016}
R.~Frank, M.~Macdonald, and B.~Monk.
\newblock Location, location, location: Mapping potential canadian targets in
  online hacker discussion forums.
\newblock EISIC '16, 2016.

\bibitem{Joobin2017}
J.~Gharibshah, T.~C. Li, M.~S. Vanrell, A.~Castro, K.~Pelechrinis, E.~E.
  Papalexakis, and M.~Faloutsos.
\newblock Inferip: Extracting actionable information from security discussion
  forums.
\newblock In {\em Proceedings of the 2017 IEEE/ACM International Conference on
  Advances in Social Networks Analysis and Mining 2017}, ASONAM '17, pages
  301--304, New York, NY, USA, 2017. ACM.

\bibitem{Hang2016}
H.~Hang, A.~Bashir, M.~Faloutsos, C.~Faloutsos, and T.~Dumitras.
\newblock {``Infect-me-not"}: A user-centric and site-centric study of
  web-based malware.
\newblock In {\em IFIP Networking}, pages 234--242, May 2016.

\bibitem{Iannacone2015}
M.~Iannacone, S.~Bohn, G.~Nakamura, J.~Gerth, K.~Huffer, R.~Bridges,
  E.~Ferragut, and J.~Goodall.
\newblock Developing an ontology for cyber security knowledge graphs.
\newblock In {\em Proceedings of the 10th Annual Cyber and Information Security
  Research Conference}, CISR '15, pages 12:1--12:4, New York, NY, USA, 2015.
  ACM.

\bibitem{mike2017}
T.~C. Li, J.~Gharibshah, E.~E. Papalexakis, and M.~Faloutsos.
\newblock Trollspot: Detecting misbehavior in commenting platforms.
\newblock In {\em Proceedings of the 2017 IEEE/ACM International Conference on
  Advances in Social Networks Analysis and Mining 2017}, ASONAM '17, 2017.

\bibitem{Motoyama:2011}
M.~Motoyama, D.~McCoy, K.~Levchenko, S.~Savage, and G.~M. Voelker.
\newblock An analysis of underground forums.
\newblock In {\em Proceedings of the 2011 ACM SIGCOMM Conference on Internet
  Measurement Conference}, IMC '11, pages 71--80, New York, NY, USA, 2011. ACM.

\bibitem{papalexakis2013k}
E.~E. Papalexakis, N.~D. Sidiropoulos, and R.~Bro.
\newblock From k-means to higher-way co-clustering: Multilinear decomposition
  with sparse latent factors.
\newblock {\em IEEE transactions on signal processing}, 61(2):493--506, 2013.

\bibitem{Portnoff2017}
R.~S. Portnoff, S.~Afroz, G.~Durrett, J.~K. Kummerfeld, T.~Berg-Kirkpatrick,
  D.~McCoy, K.~Levchenko, and V.~Paxson.
\newblock Tools for automated analysis of cybercriminal markets.
\newblock WWW '17, 2017.

\bibitem{Ramos2003UsingTT}
J.~Ramos.
\newblock Using {TF-IDF} to determine word relevance in document queries.
\newblock In {\em {Instructional Conference on Machine Learning}}, 2003.

\bibitem{Reza2016}
R.~Rawassizadeh, E.~Momeni, C.~Dobbins, J.~Gharibshah, and M.~Pazzani.
\newblock Scalable daily human behavioral pattern mining from multivariate
  temporal data.
\newblock {\em IEEE Transactions on Knowledge and Data Engineering},
  28(11):3098--3112, Nov 2016.

\bibitem{Az1}
S.~Samtani, R.~Chinn, and H.~Chen.
\newblock Exploring hacker assets in underground forums.
\newblock In {\em IEEE International Conference on Intelligence and Security
  Informatics (ISI)}, pages 31--36, May 2015.

\bibitem{Johan2011}
J.~Ugander, B.~Karrer, L.~Backstrom, and C.~Marlow.
\newblock The anatomy of the facebook social graph.
\newblock {\em CoRR}, abs/1111.4503, 2011.

\bibitem{Yang:1997:CSF:645526.657137}
Y.~Yang and J.~O. Pedersen.
\newblock A comparative study on feature selection in text categorization.
\newblock In {\em Proceedings of the Fourteenth International Conference on
  Machine Learning}, ICML '97, pages 412--420, San Francisco, CA, USA, 1997.
  Morgan Kaufmann Publishers Inc.

\bibitem{Zhang:2015}
X.~Zhang, A.~Tsang, W.~T. Yue, and M.~Chau.
\newblock The classification of hackers by knowledge exchange behaviors.
\newblock {\em Information Systems Frontiers}, 17(6):1239--1251, Dec. 2015.

\end{thebibliography}
%\bibliography{BIB/satc.bib}

%\balance

%\clearpage
%\input{appendix.tex}

\end{document}